\documentclass[draft, english]{volcanica-template} 

\newcommand{\Title}{Cavity enhancement of V2 centers in 4H-SiC with a fiber-based Fabry-Pérot microcavity} 
\newcommand{\shortTitle}{Cavity enhancement of V2 centers in 4H-SiC} 
\newcommand{\Author}{J. Hessenauer et al.\xspace} 
\author[{{\affiliation{1}}}]
{Jannis Hessenauer~\orcidaffil{0000-0001-9586-7878}\Email{jannis.hessenauer@kit.edu}\SharedAuthor}

\author[{{\affiliation{2}}}]
{Jonathan K\"orber~\orcidaffil{0000-0002-7531-0295}\Email{jonathan.koerber@pi3.uni-stuttgart.de}\SharedAuthor}

\author[{{\affiliation{3}}}]
{Misagh Ghezellou~\orcidaffil{0000-0002-1969-3324}}

\author[{{\affiliation{3}}}]
{Jawad Ul-Hassan~\orcidaffil{0000-0001-9537-2226}}

\author[{{\affiliation{4}}}]
{Georgy V. Astakhov~\orcidaffil{0000-0003-1807-3534}}

\author[{{\affiliation{5}}}]
{Wolfgang Knolle}

\author[{{\affiliation{2},\affiliation{6}}}]
{Jörg Wrachtrup~\orcidaffil{0000-0003-3328-9093}}

\author[{{\affiliation{1},\affiliation{7}}}]
{David Hunger~\orcidaffil{0000-0001-6156-6145}}

\affil[{{\affiliation{1}}}]{					
Physikalisches Institut, Karlsruhe Institute of Technology (KIT), Wolfgang-Gaede-Str. 1, 76131 Karlsruhe, Germany.
}

\affil[{{\affiliation{2}}}]{					
3rd Institute of Physics, University of Stuttgart, Allmandring 13, 70569 Stuttgart, Germany.
}

\affil[{{\affiliation{3}}}]{					
Department of Physics, Chemistry and Biology, Linköping University, 581 83 Linköping, Sweden.
}

\affil[{{\affiliation{4}}}]{					
Institute of Ion Beam Physics and Materials Research, Helmholtz-Zentrum Dresden-Rossendorf, 01328 Dresden, Germany.}

\affil[{{\affiliation{5}}}]{					
Leibniz-Institute of Surface Engineering (IOM), Permoserstraße 15, 04318 Leipzig, Germany.
}

\affil[{{\affiliation{6}}}]{                    
Max Planck Institute for Solid State Research, Heisenbergstraße 1, 70569 Stuttgart, Germany.}

\affil[{{\affiliation{7}}}]{                    
Institute for Quantum Materials and Technologies (IQMT), Karlsruhe Institute of Technology (KIT), Herrmann-von-Helmholtz Platz 1, 76344 Eggenstein-Leopoldshafen, Germany.}

\begin{document}


\FrontMatter{\protect{
\noindent Silicon vacancy centers in 4H-silicon carbide (SiC) host a long-lived electronic spin and simultaneously possess spin-resolved optical transitions, making them a great candidate for implementing a spin-photon interface. These interfaces are an important building block of quantum networks, which in turn promise to enable secure communication and distributed quantum computing. To achieve this goal, the rate of coherently scattered photons collected from a single emitter needs to be maximized, which can be achieved by interfacing the emitter with an optical cavity. In this work, we integrate V2 centers inside a SiC membrane into a fiber-based Fabry-P\'erot microcavity. We find that SiC lends itself well to membrane fabrication, as evidenced by low surface roughness $\sigma_\mathrm{RMS} \approx \SI{400}{pm}$, high reproducibility, and consequentially a high cavity finesse $\mathcal{F} \approx \SI{40000}{}$. At cryogenic temperatures, we observe individual emitters coupling to cavity modes. We confirm their single-emitter character by measuring the second-order autocorrelation function and investigate the Purcell factor by measuring the optical lifetime as a function of the cavity-emitter detuning. We find a 13.3-fold enhancement of photons scattered into the zero-phonon line (ZPL), which could be further increased by using optimized mirror coatings, potentially opening the path towards deterministic spin-photon interaction. 

}}[]{}

\section*{INTRODUCTION}\label{sec:introduction}
Spin-active quantum emitters in solids are considered a promising platform for quantum technology applications \cite{Atatuere2018,Awschalom2018}. Diamond, as the main host material for such emitters, has been used for pioneering experiments in the field \cite{Gaebel2006,Balasubramanian2008,Neumann2010,Hensen2015,Bradley2019,Pompili2021}. However, other semiconductor materials such as silicon (Si) \cite{Durand2021,Higginbottom2022, Hollenbach2022} and SiC \cite{Castelletto2020,Castelletto2022} have recently gained increasing attention because they offer advantages such as CMOS-fabrication compatibility and wafer-scale material availability \cite{Ryu1998,Hassan2008,Liu2015,Yang2022} while still hosting promising quantum emitters, such as silicon-vacancy centers (V$_{\mathrm{Si}}$) \cite{Kraus2013,Widmann2014,Nagy2019} or divacancy centers (VV) \cite{Son2006,Christle2017,Li2022} in 4H-SiC. The V$_{\mathrm{Si}}$-center in 4H-SiC has already been used for encouraging demonstrations of single-shot readout of the spin state for the k-site V$_{\mathrm{Si}}$ (V2) \cite{Lai2024,Hesselmeier2024} and spin-photon entanglement for the h-site V$_{\mathrm{Si}}$ (V1) \cite{Fang2024}. However, it shows only low Debye-Waller factors (\SIrange{6}{9}{\percent}) and a moderate quantum efficiency of QE=\SI{28.6}{\percent}, thus raising challenges for useful quantum technology applications with the system \cite{Udvarhelyi2020,Son2020,Shang2020,Liu2024}.

Different photonic structures have already been demonstrated to enhance the collected photon rate \cite{Radulaski2017,Sardi2020,Bekker2023,Zhou2023, Krumrein2024, Koerber2024}. Yet, to specifically enhance the ZPL transition of the emitters, highly resonant structures that can harness the Purcell effect must be used. This increases the rate of coherent photons, which are necessary to generate and distribute remote entanglement in quantum networks \cite{Awschalom2018}. Photonic crystal cavities and disk resonators have already been used to enhance V$_{\mathrm{Si}}$- and VV-centers in 4H-SiC \cite{Crook2020,Lukin2020,Lukin2023}. Compared to these integrated structures, open microcavities such as (fiber-based) Fabry-Pérot cavities offer the possibility of easy-tunability of the spatial and spectral position of the cavity resonance \cite{Hunger2010,Toninelli2010,Mader2015,Fait2021}. This enables the in-situ optimization of resonance conditions and the investigation of multiple emitters with a single cavity. With this approach, emitters can be integrated within thin membranes bonded onto the cavity mirror, as already successfully demonstrated for the diamond platform \cite{Janitz2015, Riedel2017, Heupel2020, Salz2020, Ruf2021,Pallmann2024}.
Crucially, this allows one to position emitters farther from surfaces, avoiding spectral diffusion due to charge fluctuations. 

In this work, we demonstrate the coupling of V2 centers inside a few-micrometer-thin 4H-SiC membrane to a fiber-based Fabry-Pérot cavity. We produce two membranes with an extremely low surface roughness of \SIrange{350}{400}{\pico\meter} RMS and high spatial homogeneity. We study the losses and dispersion of the cavity-membrane system and find excellent agreement with theoretical descriptions and simulations, as well as minimal loss introduced by the membrane, for both samples. We then analyze the emitter properties and find a reduced lifetime when resonant with the cavity due to the Purcell effect. Finally, we show that by using the cavity to spectrally select individual emitters, we achieve high single-photon purity and high ZPL count rates.

\section*{RESULTS AND DISCUSSION}\label{sec:results}
\subsection*{Sample fabrication and characterization}
\begin{figure}[tb]
\includegraphics[width=\linewidth]{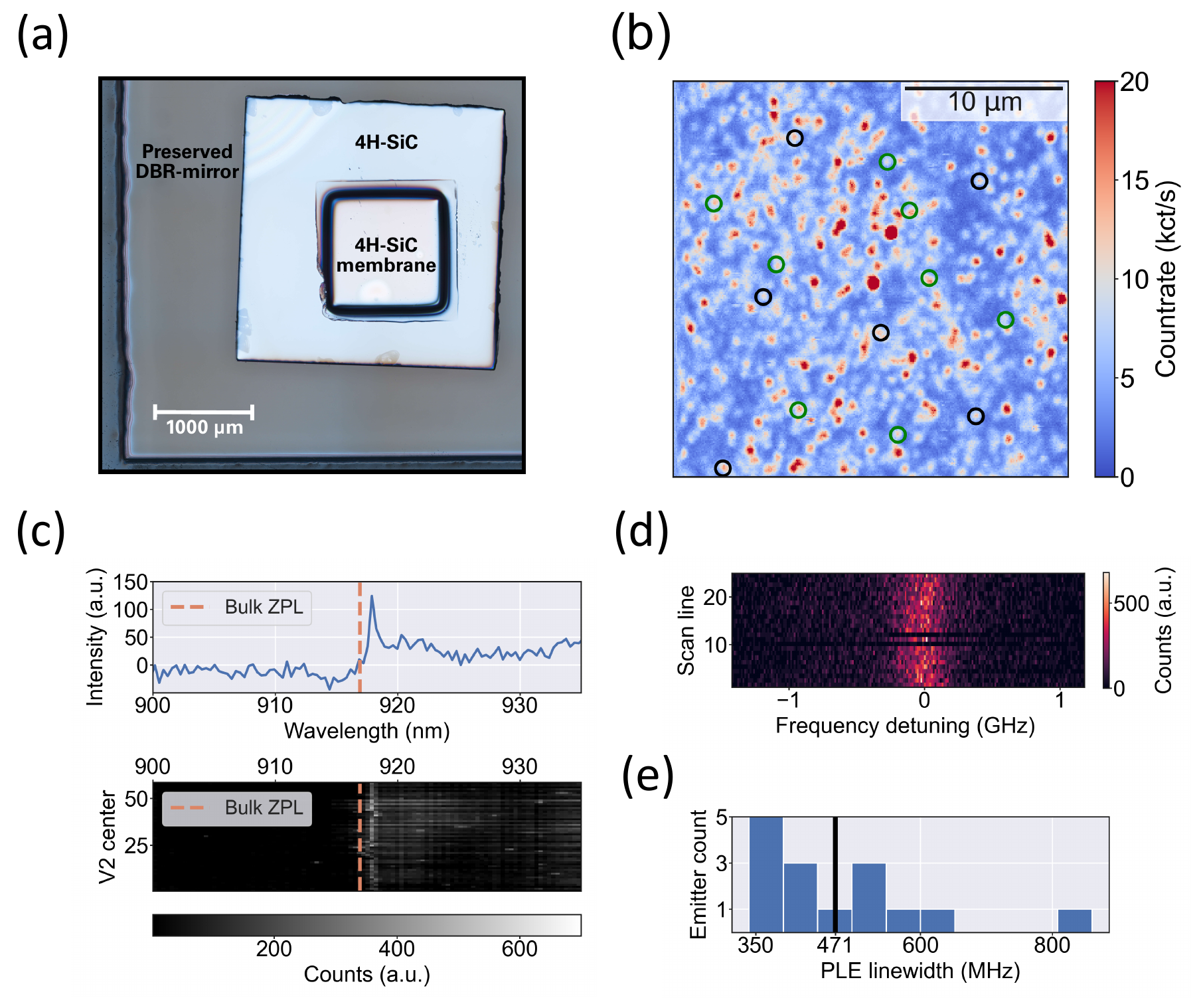}
\centering
\caption[]{\textbf{Fabrication and characterization of the membrane SA.}
(a) Light microscope image of the sample after the fabrication. (b) Room temperature confocal fluorescence scan in the membrane region of the sample. Diffraction limited spots that were further investigated by ODMR are marked with a circle. The green circles show spots with a clear ODMR signature from a V2 center. (c) Background corrected emission spectrum of a single spot (upper panel) under off-resonant excitation and integration for 20 seconds at a temperature of \SI{8}{\kelvin} and spectra of 58 investigated V2 centers stacked above each other (lower panel). The orange, dashed line indicates the ZPL position from bulk V2 centers. (d) PLE scans of a single V2 center in the membrane at \SI{8}{\kelvin}. The two dark lines are due to an ionization of the emitter upon resonant excitation. (e) PLE-linewidth distribution of different V2 centers in the membrane. The linewidth is extracted by the average FWHM of Lorentzian fits on 25 single-line PLE scans for each center. The black solid line at \SI{471}{\mega\hertz} depicts the average linewidth  of all investigated V2 centers. 
}
\label{Fig1}
\end{figure}
The samples we use for our experiments are made from a commercial, n-type, a-plane 4H-SiC wafer (\textit{Wolfspeed}) with a 10-micron-thick high quality epilayer grown on the top side. In total, we fabricate and investigate two samples, sample A (SA) and sample B (SB). Analogously to previous work \cite{Heiler2024,Koerber2024}, we use a series of lapping and chemical-mechanical polishing (CMP) to thin the samples from the wafer side down to a total thickness of \SI{\sim 40}{\micro\meter}. Subsequently, the samples are van der Waals bonded with the epilayer side onto a commercially coated distributed-bragg-reflector (DBR)-mirror (\textit{Laseroptik}), using an oxygen plasma activation (see supplement for details on the bonding procedure). Next, we thin the samples on the mirror further down to a few \SI{}{\micro\meter}-thin membranes using a SF$_6$-based reactive-ion etching (RIE) process. In order to prevent a strong amount of micromasking due to nonvolatile etch by-products from the mirror and to preserve parts of the free mirror throughout the etching, we use a SiC hard mask with a square opening on top of the silicon carbide sample SA and only etch it within the opening of the mask. This yields only a square sub-region of the whole SiC sample at the target thickness as visible in the microscope image of Figure \ref{Fig1} (a). SB is etched without a hard mask on top. Our etching leaves the surface with a low RMS roughness of \SI{\sim 400}{\pico\meter} (see supplement for AFM measurements), which is crucial for high-finesse cavity experiments. To create V2-centers, we use electron irradiation at an energy of \SI{5}{\mega\electronvolt} and a dose of \SI{5}{\kilo\gray} (see supplement for more information on the electron irradiation and for a full process flow diagram of the fabrication). 

To confirm the presence of V2 centers in the fabricated samples, we use a home-built confocal microscope at room temperature (NA=0.9 objective). Figure \ref{Fig1} (b) shows a typical fluorescence map of a central region within the membrane SA under off-resonant excitation at \SI{785}{\nano\meter}. Since the V2 emission spectrum has no characteristic peak at room temperature, we use optically detected magnetic resonance (ODMR) measurements to identify V2 centers together with a preselection based on the polarization of the fluorescence (see \cite{Koerber2024} for details on the preselection). Out of 14 preselected spots, we find 8 that show a clear ODMR signal from a V2 center (see supplement for details on the ODMR measurements). The remaining six spots are most likely V1 centers.

\begin{figure*}[tb]
    \centering    \includegraphics[width=0.9\linewidth]{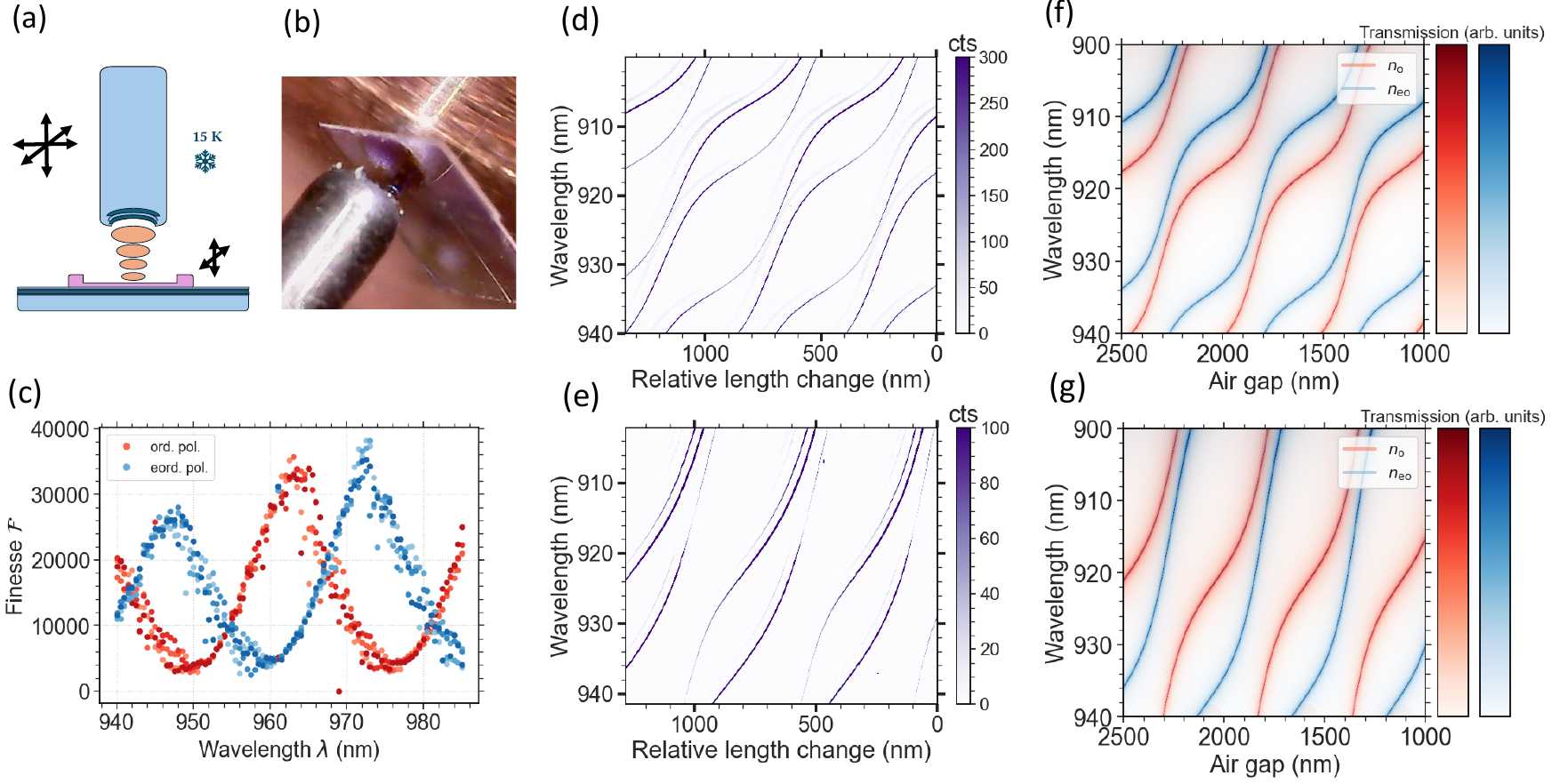}    \caption{\textbf{Characterisation of the membrane cavity system.} (a) Schematic sketch of the fiber-cavity-membrane system. (b) Photograph of the fiber mirror protruding from a metallic needle opposing the membrane bonded onto a planar DBR  mirror. (c) Finesse as a function of wavelength for two orthogonal polarisations, corresponding to the ordinary and extraordinary polarization modes of the SiC membrane. The periodic modulation is due to the SiC membrane. (d),(e) Measured cavity dispersion of SA and SB, showcasing the characteristic hybridisation of air-like and dielectric-like modes for membrane cavity systems. Due to the birefringence of SiC, we observe two families of modes. The different membrane thickness changes the periodicity of the mode coupling. The fainter, slightly offset traces correspond to higher order cavity modes. (f),(g) Simulated dispersion for SA,SB. The simulation reproduces the measurements. Extracted membrane thicknesses are $\mathrm{d_{SA}} = \SI{6.20}{\micro m}$ and $\mathrm{d_{SB}} = \SI{2.85}{\micro m}$.}
    \label{fig:Cavity_Characterization}
\end{figure*}

To characterize the optical properties of the V2 centers in SA, we cool the sample to \SI{8}{\kelvin} in a closed-cycle cryostat (\textit{Montana Instruments}), and use a second home-built confocal microscope (NA=0.75 objective) for cryogenic spectroscopy. Here, we identify V2 centers from their emission spectrum under off-resonant excitation at \SI{730}{\nano\meter}, as shown for an exemplary emitter in the upper panel of Figure \ref{Fig1} (c). Compared to the ZPL of the bulk centers at \SI{917}{\nano\meter} \cite{Wagner2000,Udvarhelyi2020}, we find that the ZPLs of most emitters in the bonded membrane are shifted by \SI{\sim 1}{\nano\meter}, as depicted in the stacked spectra plot of 58 different emitters in the lower panel of Figure \ref{Fig1} (c). We attribute this shift to a homogeneous strain inside the sample arising from the bonding to the DBR-mirror: A ZPL shift of \SI{1}{\nano\meter} could be explained by strain of \SI{0.02}{\percent} parallel to the c-axis \cite{Udvarhelyi2020}, which is a realistic amount since similar strain has been already observed for bonded SiC as a consequence of lattice mismatch \cite{Breev2021}.

Finally, we investigate the linewidths of the optical transitions of membrane-integrated V2 centers by photoluminescence excitation (PLE) scans. As described in earlier work \cite{
Koerber2024}, we use acousto-optic modulators (AOMs) to generate a laser beam containing two wavelengths split by \SI{1}{\giga\hertz}. With this beam, we simultaneously scan over both optical transitions of the V2 center at a power of \SI{5}{\nano\watt} before the objective and collect the photons from the phonon sideband (PSB). Figure \ref{Fig1} (d) shows a series of such scans from the same emitter stacked over each other with no visible spectral jumps. To evaluate the linewidth, we fit a Lorentzian to each individual line and extract the mean linewidth from the series. We repeat these measurements for a total of 15 emitters and find a mean PLE linewidth of \SI{471\pm 132}{\mega\hertz} for all emitters, as depicted in Figure \ref{Fig1} (e). To exclude a significant power broadening of the optical linewidths, we perform excitation power-dependent PLE linewidth measurements on a single emitter and find that there is no significant power broadening at \SI{5}{\nano\watt} excitation power (see supplement). We note that the average PLE linewidth of \SI{471\pm 132}{\mega\hertz} appears to be higher than the linewidth reported from similar samples \cite{Heiler2024,Koerber2024}. Possible explanations for the increased linewidth include non-perfect thermalization of the sample and an increased amount of charge noise, both stemming from the thick, isolating glass substrate of the mirror between the silicon carbide membrane and the copper plate inside the cryostat.

\subsection*{Cavity characterization}
In order to characterize the membrane-mirror system performance, we integrate the sample into a home-built positioning unit, forming a microcavity together with a fiber mirror that is approached closely to the membrane, as sketched in Figure \ref{fig:Cavity_Characterization} (a) and depicted in Figure \ref{fig:Cavity_Characterization} (b). The positioning setup is located in a closed-cycle cryostat as reported in earlier work \cite{Pallmann2023}. The fiber mirror can be scanned laterally across the mirror over a range of $\SI{20}{\micro \meter} \times \SI{20}{\micro \meter}$ and the cavity length can be tuned over multiple free spectral ranges. This enables us to spatially search for emitters, optimize the lateral overlap of emitter and cavity mode and then minimize the cavity length by bringing the fiber mirror in contact with the membrane. 
The contact configuration has the added benefit of drastically improving the longitudinal stability. This allows us to avoid active stabilization and passively stay on resonance with the ZPL despite a relatively high finesse. This configuration is used for all fluorescence measurements in this work (data presented in Figure \ref{fig:emitter-dispersion} and \ref{fig:cavity-fluorescence-analysis}).

We begin by measuring the finesse at small mirror distance by monitoring the transmission of a narrow linewidth laser through the cavity, while modulating the cavity length over multiple free spectral ranges. The transmission spectrum is fitted with Lorentzians to determine the positions and linewidths of the fundamental modes, allowing us to calculate the finesse. The finesse as a function of the probe wavelength is displayed in Figure \ref{fig:Cavity_Characterization} (c).
It exhibits a strong dependence on the probe wavelength, due to the modulation of the DBR transmission by the membrane. The presence of a membrane in the cavity leads to the formation of a dielectric and air-like set of modes; see \cite{VanDam2018OptimalApproach, Koerber2023} and below. We observe a finesse as high as $\mathcal{F}=40000$ at a wavelength of $\lambda = \SI{980}{ \nano \metre}$, which is almost twice the value observed for a bare cavity. From simulations with a transfer matrix model, we can extract additional losses due to scattering and absorption in the high-finesse case of approximately $\mathcal{L}_{\mathrm{air}} \approx 70\,\mathrm{ppm}$, further evidencing the high sample quality. We measure a finesse of up to $\mathcal{F}=10000$ at $\lambda = \SI{917}{nm}$, where the zero-phonon line of the V2 center is located.

Next, we measure the dispersion of the cavity-membrane system by coupling a broadband light source to the cavity and observing the resonances in transmission with a spectrometer, while stepwise changing the cavity length (Figure \ref{fig:Cavity_Characterization} (d) and (e)). The mode hybridization of the cavity with the membrane manifests itself in a modulation of the cavity dispersion, as previously observed in comparable systems \cite{Jensen2020,Salz2020,Janitz2015,Koerber2023,Maisch2024}. It is further complicated by the strong birefringence of 4H-SiC, which fully determines the polarization modes of the cavity and leads to two families of modes with orthogonal polarization \cite{Xing2019}. The two polarization modes are also observed in the finesse measurement (Figure \ref{fig:Cavity_Characterization} (c)). 

Simulated dispersion plots of the cavity membrane system using a transfer matrix model \cite{Janitz2015} are shown in Figure \ref{fig:Cavity_Characterization} (f),(g)). The good agreement with the experimental data confirms that our model captures the relevant effects and allows us to extract the membrane thickness of SA to be $d_\mathrm{SA} = \SI{6.2}{\micro m}$ and of SB to be $d_\mathrm{SB} = \SI{2.85}{\micro m}$, in agreement with the measurements during fabrication using a profilometer.
 
\subsection*{Cavity enhanced V2 emission}
\begin{figure}
    \centering    \includegraphics[width=\linewidth]{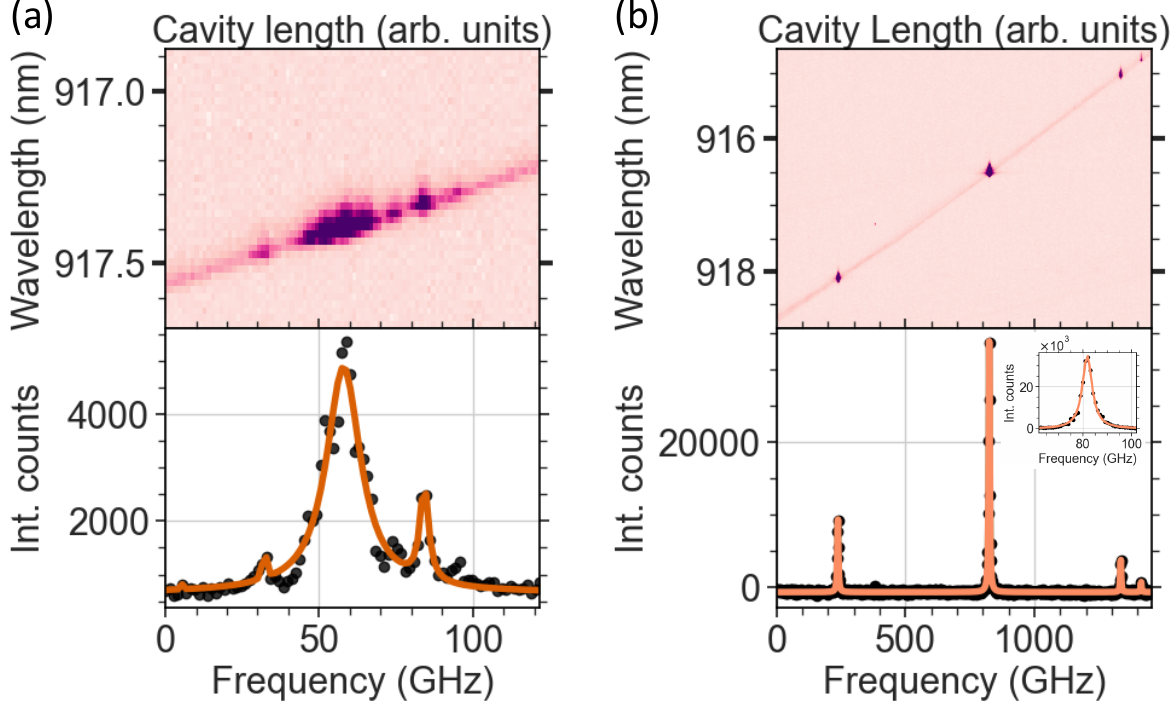}
    \caption{\textbf{Coupling of ZPLs to cavity modes} Cavity dispersion measured under off-resonant excitation. A strong increase in emission is observed when the cavity is resonant with the ZPLs (top panels). By integrating the individual spectra and calibrating the local dispersion, we can extract a frequency distribution of ZPLs (bottom panels). The orange line is a fit of multiple Lorentzians to the data. Notably, the thicker membrane SA (a) shows a denser and more centered distribution of emitters, while SB (b) exhibits sharp peaks with large spacings.} 
    \label{fig:emitter-dispersion}
\end{figure}
We continue by measuring the properties of the emitters when coupled to the cavity. We couple an off-resonant diode laser ($\lambda = 785\,\mathrm{nm}$) to the cavity fiber to excite the emitters via the PSB. The mirror coating is almost transmissive at this wavelength, but still leads to a modulation of the intensity by a factor of $\approx 3$, which we account for by measuring the excitation laser power after the cavity. We use filters to block the excitation light and detect the fluorescence in a spectral band of $\lambda = 900 \, \mathrm{nm} - 1000 \, \mathrm{nm}$, either using a spectrometer or two single-photon counters arranged in a Hanbury-Brown-Twiss (HBT) configuration.

First, we record spectra under off-resonant excitation while slowly varying the cavity length. Due to a broad background, we observe the cavity dispersion as in Figure \ref{fig:Cavity_Characterization} (d-e) as well as a strong increase in fluorescence when the cavity is resonant with the ZPL of one or more V2 centers (Figure \ref{fig:emitter-dispersion}). Crucially, this is only observed for one polarization branch, because the dipole axis of the V2 center is oriented along the crystallographic c-axis, as is the extraordinary axis of the refractive index ellipsoid (see supplement). This is useful for our application because it guarantees a maximal overlap between the emitter dipole and the cavity polarization, which otherwise would reduce the Purcell factor.

\begin{figure*}
    \centering    \includegraphics[width=0.84\textwidth]{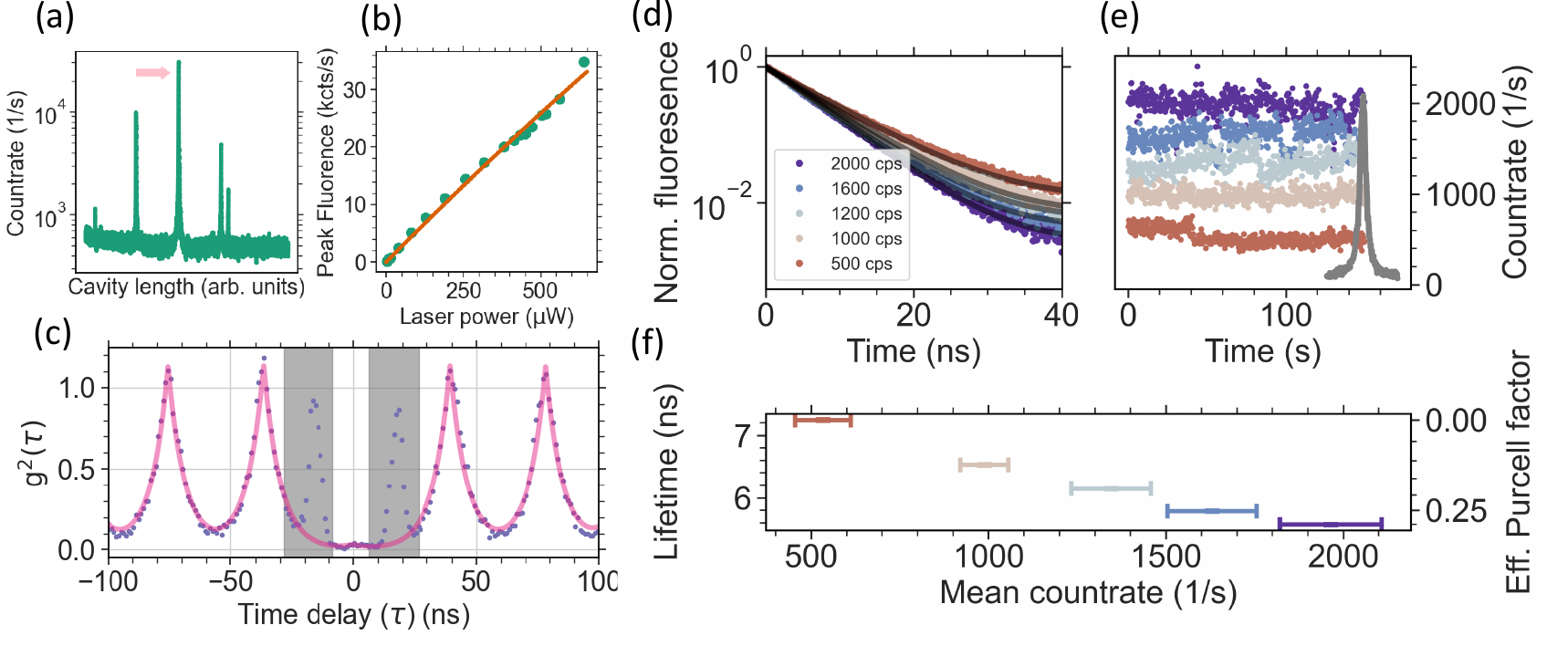} \caption{\textbf{Characterization of the emitter fluorescence in the cavity.} (a) Detected countrate under off-resonant, pulsed excitation as a function of cavity detuning. At least four distinct peaks belonging to different ZPLs are visible. The brightest peak shows a ZPL count rate of $\SI{30}{kcps}$. Note the close similarity to the measurement in Figure \ref{fig:emitter-dispersion} (b), when accounting for the different y-axis scaling. (b) Excitation power dependence of the ZPL count rate under pulsed excitation. The laser power was measured behind the cavity. No clear saturation behaviour was observed. (c) Pulsed auto-correlation measurement with the cavity resonant with the peak marked in (a), showcasing a high single-photon purity of $g^{(2)}(0) =\SI{0.024 (0.024)}{}$ without background correction. Coincidence counts in the gray area arise from recombination photons in the counting APDs and are excluded from the fit (see supplement for details). (d) Detuning dependent optical lifetime measurements of the resonance marked in (a). The detuning is reflected by the average countrate during the measurement. Solid lines show monoexponential fits with a constant offset to account for background counts. The countrate traces recorded during the lifetime measurement are plotted together with a sweep over the resonance in (e), in order to illustrate the cavity-emitter detuning and its stability. The lifetime reduces as the cavity becomes resonant with the ZPL, consistent with Purcell enhancement. The effective Purcell factors calculated from $C =\frac{\tau_0}{\tau}-1$ are plotted in (f).}
    \label{fig:cavity-fluorescence-analysis}
\end{figure*}

By performing these scans slowly and finely resolved, we notice a substructure around the ZPL of SA (Figure \ref{fig:emitter-dispersion} (a)), which we attribute  to the inhomogeneous linewidth, originating from the different local electrical environment of multiple emitters. In our sample, multiple emitters are located inside the cavity mode waist, with the depth being randomly distributed. Thus, the lateral overlap and the position in the standing wave field, and therefore the coupling to the cavity mode and the observed peak count rate, vary for the different emitters.
Interestingly, emitters in SB are both more sparse and spread over a larger frequency space (Figure \ref{fig:emitter-dispersion} (b)). Both facts can possibly be explained by the thinner membrane.

By fitting a linear slope to the local dispersion, we can calibrate the cavity length shift in terms of resonance frequency shift, allowing us to determine the feature linewidth by fitting the calibrated integrated spectrum. The observed linewidth is a convolution of the linewidths of the cavity and the emitter.  For sample SB, the brightest peak yields a linewidth of $\delta \nu = \SI{4.8}{GHz}$, while the narrowest linewidth is obtained for the two rightmost peaks with $\delta \nu = \SI{3.8}{GHz}$ (lower panel of Figure \ref{fig:emitter-dispersion} (b)). We measure the cavity linewidth at this position to be  $\delta \nu_\mathrm{cav} = \SI{3.44}{GHz}$ by slowly (\SI{0.1}{Hz}) scanning a resonant laser over the cavity and recording the frequency with a wavemeter (see supplement). This indicates a contribution from the emitter linewidth for the most dominant peak, which could be caused by insufficient thermalization of the mirror in the cavity stage. The temperature we measure at the mirror holder is \SI{16}{K}. The linewidth of V2 centers is known to significantly increase above \SI{20}{K} \cite{Udvarhelyi2020}. However, the sample is isolated from the copper mirror holder by the mirror substrate. To alleviate this, we add a copper link connecting the mirror holder to the sample, but a temperature gradient might nevertheless still persist.   

Even though multiple emitters are located inside the cavity mode volume, we can make use of their inhomogeneous spectral distribution together with the spectral selectivity of the cavity to select individual emitters. In the thinner membrane SB, we observe peaks that are well isolated and are spread over a wide spectral range (\SIrange{918}{923}{nm}). We set the cavity resonant with the brightest peak and measure the second-order autocorrelation $g^{(2)}(\tau)$ under pulsed off-resonant excitation (SuperK-Fianium, repetition rate $\SI{23.4}{MHz}$, $\lambda = \SI{780}{nm}- \SI{790}{nm} $, $P= \SI{230}{\micro\watt}$). The measurement and a fit are displayed in Figure \ref{fig:cavity-fluorescence-analysis} (c). We observe a strong antibunching dip at $\tau = \SI{0}{ns}$ with $g^{(2)}(0) = \SI{0.024 \pm 0.024}{}$, indicating a high single-photon purity. This underlines the ability of the cavity to spectrally select single emitters even under non-selective excitation, potentially enabling spectral multiplexing.

In order to quantify the achieved Purcell enhancement, we conduct lifetime measurements using the same pulsed laser source as above by recording the start-stop histograms between the laser emission and the detected photons. We vary the cavity emitter detuning and observe a lifetime-shortening as the cavity becomes resonant with the ZPL (see Figure \ref{fig:cavity-fluorescence-analysis} (d-f)). The shortest lifetime observed when the cavity and the emitter line coincide is $\tau_\mathrm{min} = \SI{5.6}{ns}$. We take the longest lifetime observed for large detunings as the free space lifetime $\tau = \SI{7.3}{ns}$, in good agreement with the values reported in the literature, e.g. $\tau = \SI{7.08}{ns}$ in \cite{Babin2022}. The lifetime shortening allows us to determine the effective Purcell factor defined as $C_\mathrm{eff} = \frac{\tau_0}{\tau_\mathrm{cav}}-1 = \SI{0.30 \pm 0.01}{}$. The observed change in the lifetime is small because only the photons emitted into the narrow ZPL experience a significant Purcell enhancement, while all other decay channels remain unperturbed. Taking into account the fraction of coherently emitted photons via the Debye-Waller factor of $\mathrm{DWF = \SI{8}{\%}}$ \cite{Udvarhelyi2020} and accounting for the quantum efficiency of $\mathrm{QE = \SI{28.6}{\%}}$ \cite{Liu2024}
, this corresponds to an ideal Purcell factor $C_0 = C_\mathrm{eff}/(\mathrm{DWF} \cdot\mathrm{QE}) = 13.3 $. The ideal Purcell factor quantifies how many ZPL photons are additionally emitted into the cavity mode compared to free-space emission into $4\pi$.
\subsection*{Discussion}
For calculating the expected Purcell factor, we use the quality factor $Q = \SI{7.4e4}{}$ extracted from the dispersion measurement (Fig \ref{fig:emitter-dispersion} (b) inset), which is sensitive to both the linewidth of the cavity and the emitter. The mode volume $\SI{12.0}{\lambda_{\mathrm{V2}}^3}$ is calculated by numerically integrating the field distribution obtained by the 1D-simulation, yielding the effective cavity length  $L_{\mathrm{eff}} = \frac{\int |n(z)E(z)|^2 dz} {\max(n(z)E(z))}$.
The mode volume is then given by $V_\mathrm{} = \frac{\pi w_\mathrm{0}^2L_\mathrm{eff}}{4}$, where the mode waist $w_0 = \SI{1.66}{\micro\meter}$ in the membrane is calculated as previously reported in \cite{Janitz2015,VanDam2018OptimalApproach} by matching a Gaussian beam in the air gap to one in the membrane. This results in an expected Purcell factor of $C_0 = \SI{25.7}{}$. By using the quality factor obtained in the dispersion measurement, we automatically account for additional broadening of the emitters in the calculation, which could be caused by insufficient thermalization.
Yet, the calculated value is almost twice as high as the observed Purcell enhancement. We believe that this difference can be explained by a non-perfect overlap of the defect with the cavity mode field, both transversally and longitudinally. Further, our calculations of the ideal Purcell factor depend on literature values for the Debye-Waller factor and the quantum efficiency. While these properties have been studied carefully in the past, it is unclear how they are affected by a thin membrane, in particular because of applied stress. 

However, even a relatively small lifetime reduction due to the small QE and DWF leads to a large increase of ZPL photons by a factor of $C_0$, which is confirmed by the observed ZPL count rates of up to $\SI{30}{kcts/s}$ (Figure \ref{fig:cavity-fluorescence-analysis} (a)). We note that we use conventional silicon-based single photon counters, which feature a smaller detection efficiency (detection efficiency $\approx \SI{30}{\%}$) than the commonly used superconducting nanowire detectors (detection efficiency $\approx \SI{90}{\%}$).
Furthermore, due to the limited available pulsed laser power and inefficient excitation due to the large cavity mode waist, we are unable to saturate the emitter. This is evident from both the saturation measurement depicted in Figure \ref{fig:cavity-fluorescence-analysis} (b) and the rather unpronounced bunching in $g^{(2)}(\tau)$ displayed in Figure \ref{fig:cavity-fluorescence-analysis} (c). We estimate that by driving the emitter in saturation, we could increase the count rate by at least a factor of four. Therefore, we predict that we can achieve more than an order of magnitude higher count rates by using a more potent laser source and state of the art single photon detectors, without any changes to the cavity-membrane system.

\subsection*{Summary and Outlook}
In conclusion, we have demonstrated Purcell enhancement of single V2 centers in a few-micron-thick 4H-SiC membrane by integrating the membrane in a fiber-based Fabry-Pérot cavity. The suitability of SiC for this approach is evident by the reproducible fabrication process, excellent surface roughness, and experimentally observed high finesse. We find a lifetime reduction of $C_\mathrm{eff} = \SI{0.30}{}$, which corresponds to a 13.3-fold enhancement of the coherent zero-phonon line. The spectral selectivity of the cavity allows us to achieve a high single-photon purity even though multiple defects are spatially located inside the cavity field. We detect up to \SI{30}{kcts \per s} ZPL photons from a single emitter, which could be increased by more than an order of magnitude by updating the experimental setup with state of the art technology.

In future work, the Purcell factor could be increased by utilizing an optimized mirror coating with the stopband centered around the ZPL resonance. Operating an experiment with a finesse of $\mathcal{F} = \SI{40000}{}$, as measured in the center of the stopband for our mirror, requires extremely low cavity fluctuations, which, however, have been demonstrated for the setup used in this work using active stabilization. This would enable ten times higher effective Purcell factors of up to $C_\mathrm{eff} = \SI{3}{}$. For spectrally narrow emitters close to the lifetime limit, as already demonstrated for V2 centers in even thinner membranes \cite{Heiler2024}, this corresponds to a cooperativity $C>1$, enabling spin-based reflection schemes which can be harnessed to generate remote entanglement \cite{Nemoto2014}. 
Another important step forward is to demonstrate control of the spin state, either by integrating an RF line into the setup or by coherently driving the optical transitions.

\section*{DATA AVAILABILITY}
The data to reproduce the figures in the main text are available at \url{https://doi.org/10.18419/DARUS-4788}.

\section*{CODE AVAILABILITY}
The measurement and evaluation codes used for this study are available from the corresponding author upon reasonable request. 


\section*{COMPETING INTERESTS}
The authors declare no competing interests.

\section*{ACKNOWLEDGEMENTS}
We acknowledge fruitful discussions and experimental help from F. Kaiser, J. Heiler, K. Köster, S. Müller, M. Pallmann, R. Wörnle, T. Steidl, D. Liu, P. Kuna and V. Vorobyov.

J.H. and D.H. acknowledge funding from the Karlsruhe School of Optics and Photonics (KSOP).
J.H., D.H. and J.W. acknowledge funding by the German Federal Ministry of Education and Research (BMBF) within the project QR.X (Contract No. 16KISQ004).

J.W. further acknowledges funding by the German research foundation (DFG, Grant agreement No. GRK2642), the Baden-Württemberg Stiftung via the project SPOC (Grant agreement No. QT-6), as well as the BMBF via the project InQuRe (Grant agreement No. 16KIS1639K) and the Clusters4Future QSENS project QVOL. 
J.W. and J.U.H. acknowledge the European Union (EU) within the Horizon Europe project SPINUS (Grant agreement No. 101135699) and the QuantERA project InQuRe (Grant agreements No. 731473 and 101017733)

J.U.H. further acknowledges funding by the Swedish Research Council under VR Grant
No. 2020-05444.

G.A. gratefully acknowledges support from the Ion Beam Center (IBC) at HZDR for ion implantation.

\EndMatter

@article{Hunger2010,
    title = {{A fiber Fabry–Perot cavity with high finesse}},
    year = {2010},
    journal = {New J. Phys.},
    author = {Hunger, D. and Steinmetz, T. and Colombe, Y. and Deutsch, C. and H{\"{a}}nsch, T. W. and Reichel, J.},
    number = {6},
    month = {6},
    pages = {065038},
    volume = {12},
    publisher = {IOP Publishing},
    doi = {10.1088/1367-2630/12/6/065038},
    issn = {1367-2630},
    arxivId = {1005.0067}
}

@article{Pallmann2023,
    title = {{A highly stable and fully tunable open microcavity platform at cryogenic temperatures}},
    year = {2023},
    journal = {APL Photonics},
    author = {Pallmann, Maximilian and Eichhorn, Timon and Benedikter, Julia and Casabone, Bernardo and H{\"{u}}mmer, Thomas and Hunger, David},
    number = {4},
    month = {4},
    volume = {8},
    publisher = {American Institute of Physics Inc.},
    doi = {10.1063/5.0139003},
    issn = {23780967},
    arxivId = {2212.11601}
}

@article{Mader2015,
    title = {{A scanning cavity microscope}},
    year = {2015},
    journal = {Nat. Commun.},
    author = {Mader, Matthias and Reichel, Jakob and H{\"{a}}nsch, Theodor W. and Hunger, David},
    number = {1},
    month = {6},
    pages = {1--7},
    volume = {6},
    publisher = {Nature Publishing Group},
    doi = {10.1038/ncomms8249},
    issn = {2041-1723},
    arxivId = {1411.7180}
}

@article{Toninelli2010,
    title = {{A scanning microcavity for in situ control of single-molecule emission}},
    year = {2010},
    journal = {Appl. Phys. Lett.},
    author = {Toninelli, C. and Delley, Y. and St{\"{o}}ferle, T. and Renn, A. and G{\"{o}}tzinger, S. and Sandoghdar, V.},
    number = {2},
    month = {7},
    pages = {021107},
    volume = {97},
    publisher = {AIP Publishing},
    doi = {10.1063/1.3456559},
    issn = {00036951},
    arxivId = {1005.0236}
}

@article{Bradley2019,
    title = {{A Ten-Qubit Solid-State Spin Register with Quantum Memory up to One Minute}},
    year = {2019},
    journal = {Phys. Rev. X},
    author = {Bradley, C. E. and Randall, J. and Abobeih, M. H. and Berrevoets, R. C. and Degen, M. J. and Bakker, M. A. and Markham, M. and Twitchen, D. J. and Taminiau, T. H.},
    number = {3},
    month = {9},
    pages = {031045},
    volume = {9},
    publisher = {American Physical Society},
    doi = {doi.org/10.1103/PhysRevX.9.031045},
    issn = {21603308},
    arxivId = {1905.02094}
}

@article{Durand2021,
    title = {{Broad Diversity of Near-Infrared Single-Photon Emitters in Silicon}},
    year = {2021},
    journal = {Phys. Rev. Lett.},
    author = {Durand, A. and Baron, Y. and Redjem, W. and Herzig, T. and Benali, A. and Pezzagna, S. and Meijer, J. and Kuznetsov, A. Yu and G{\'{e}}rard, J. M. and Robert-Philip, I. and Abbarchi, M. and Jacques, V. and Cassabois, G. and Dr{\'{e}}au, A.},
    number = {8},
    month = {2},
    pages = {083602},
    volume = {126},
    publisher = {American Physical Society},
    doi = {10.1103/PhysRevLett.126.083602},
    issn = {10797114},
    pmid = {33709758},
    arxivId = {2010.11068}
}

@article{Jensen2020,
    title = {{Cavity-Enhanced Photon Emission from a Single Germanium-Vacancy Center in a Diamond Membrane}},
    year = {2020},
    journal = {Phys. Rev. Appl.},
    author = {H{\o}y Jensen, Rasmus and Janitz, Erika and Fontana, Yannik and He, Yi and Gobron, Olivier and Radko, Ilya P and Bhaskar, Mihir and Evans, Ruffin and Rodr{\'{i}}guez Rosenblueth, César Daniel and Childress, Lilian and Huck, Alexander and Lund Andersen, Ulrik},
    number = {6},
    month = {6},
    pages = {64016},
    volume = {13},
    doi = {10.1103/PhysRevApplied.13.064016},
    issn = {2331-7019},
    language = {en}
}

@article{Pallmann2024,
    title = {{Cavity-Mediated Collective Emission from Few Emitters in a Diamond Membrane}},
    year = {2024},
    journal = {Phys. Rev. X},
    author = {Pallmann, Maximilian and K{\"{o}}ster, Kerim and Zhang, Yuan and Heupel, Julia and Eichhorn, Timon and Popov, Cyril and M{\o}lmer, Klaus and Hunger, David},
    number = {4},
    month = {12},
    pages = {041055},
    volume = {14},
    doi = {10.1103/PhysRevX.14.041055},
    issn = {2160-3308}
}

@article{Xing2019,
    title = {{CMOS-Compatible PECVD Silicon Carbide Platform for Linear and Nonlinear Optics}},
    year = {2019},
    journal = {ACS Photonics},
    author = {Xing, Peng and Ma, Danhao and Ooi, Kelvin J.A. and Choi, Ju Won and Agarwal, Anuradha Murthy and Tan, Dawn},
    number = {5},
    month = {5},
    pages = {1162--1167},
    volume = {6},
    publisher = {American Chemical Society},
    doi = {10.1021/acsphotonics.8b01468},
    issn = {23304022}
}

@article{Widmann2014,
    title = {{Coherent control of single spins in silicon carbide at room temperature}},
    year = {2015},
    journal = {Nat. Mater.},
    author = {Widmann, Matthias and Lee, Sang Yun and Rendler, Torsten and Son, Nguyen Tien and Fedder, Helmut and Paik, Seoyoung and Yang, Li Ping and Zhao, Nan and Yang, Sen and Booker, Ian and Denisenko, Andrej and Jamali, Mohammad and Ali Momenzadeh, S. and Gerhardt, Ilja and Ohshima, Takeshi and Gali, Adam and Janz{\'{e}}n, Erik and Wrachtrup, Jörg},
    number = {2},
    month = {12},
    pages = {164--168},
    volume = {14},
    publisher = {Nature Publishing Group},
    doi = {10.1038/nmat4145},
    issn = {1476-4660}
}

@article{Salz2020,
    title = {{Cryogenic platform for coupling color centers in diamond membranes to a fiber-based microcavity}},
    year = {2020},
    journal = {Appl. Phys. B: Lasers and Optics},
    author = {Salz, M. and Herrmann, Y. and Nadarajah, A. and Stahl, A. and Hettrich, M. and Stacey, A. and Prawer, S. and Hunger, D. and Schmidt-Kaler, F.},
    number = {8},
    month = {8},
    pages = {1--13},
    volume = {126},
    publisher = {Springer},
    doi = {10.1007/s00340-020-07478-5},
    issn = {09462171},
    arxivId = {2002.08304}
}

@article{Yang2022,
    title = {{Demonstration of 4H-SiC CMOS digital IC gates based on the mainstream 6-inch wafer processing technique}},
    year = {2022},
    journal = {J. Semicond.},
    author = {Yang, Tongtong and Wang, Yan and Yue, Ruifeng and Yang, Tongtong and Wang, Yan and Yue, Ruifeng},
    number = {8},
    month = {8},
    pages = {082801--1},
    volume = {43},
    publisher = {Journal of Semiconductors},
    doi = {10.1088/1674-4926/43/8/082801},
    issn = {1674-4926}
}

@article{Riedel2017,
    title = {{Deterministic enhancement of coherent photon generation from a nitrogen-vacancy center in ultrapure diamond}},
    year = {2017},
    journal = {Phys. Rev. X},
    author = {Riedel, Daniel and S{\"{o}}llner, Immo and Shields, Brendan J. and Starosielec, Sebastian and Appel, Patrick and Neu, Elke and Maletinsky, Patrick and Warburton, Richard J.},
    number = {3},
    month = {9},
    pages = {031040},
    volume = {7},
    publisher = {American Physical Society},
    doi = {10.1103/PhysRevX.7.031040},
    issn = {21603308},
    arxivId = {1703.00815}
}

@article{Son2020,
    title = {{Developing silicon carbide for quantum spintronics}},
    year = {2020},
    journal = {Appl. Phys. Lett.},
    author = {Son, Nguyen T. and Anderson, Christopher P. and Bourassa, Alexandre and Miao, Kevin C. and Babin, Charles and Widmann, Matthias and Niethammer, Matthias and Ul Hassan, Jawad and Morioka, Naoya and Ivanov, Ivan G. and Kaiser, Florian and Wrachtrup, Jörg and Awschalom, David D.},
    number = {19},
    month = {5},
    pages = {12},
    volume = {116},
    publisher = {American Institute of Physics Inc.},
    doi = {10.1063/5.0004454},
    issn = {00036951}
}

@article{Ryu1998,
    title = {{Digital CMOS IC's in 6H-SiC operating on a 5-V power supply}},
    year = {1998},
    journal = {IEEE Trans. Electron Devices},
    author = {Ryu, Sei Hyung and Kornegay, Kevin T. and Cooper, James A. and Melloch, Michael R.},
    number = {1},
    pages = {45--53},
    volume = {45},
    doi = {10.1109/16.658810},
    issn = {00189383}
}

@article{Son2006,
    title = {{Divacancy in 4H-SiC}},
    year = {2006},
    journal = {Phys. Rev. Lett.},
    author = {Son, N. T. and Carlsson, P. and Ul Hassan, J. and Janz{\'{e}}n, E. and Umeda, T. and Isoya, J. and Gali, A. and Bockstedte, M. and Morishita, N. and Ohshima, T. and Itoh, H.},
    number = {5},
    month = {2},
    pages = {055501},
    volume = {96},
    publisher = {American Physical Society},
    doi = {10.1103/PhysRevLett.96.055501},
    issn = {10797114}
}

@article{Wagner2000,
    title = {{Electronic structure of the neutral silicon vacancy in 4⁢H and 6⁢H SiC}},
    year = {2000},
    journal = {Phys. Rev. B},
    author = {Wagner, Mt and Magnusson, B. and Chen, W. M. and Janz{\'{e}}n, E. and S{\"{o}}rman, E. and Hallin, C. and Lindstrom, J. L.},
    number = {24},
    month = {12},
    pages = {16555},
    volume = {62},
    publisher = {American Physical Society},
    doi = {10.1103/PhysRevB.62.16555},
    issn = {01631829}
}

@article{Fang2024,
    title = {{Experimental Generation of Spin-Photon Entanglement in Silicon Carbide}},
    year = {2024},
    journal = {Phys. Rev. Lett.},
    author = {Fang, Ren-Zhou and Lai, Xiao-Yi and Li, Tao and Su, Ren-Zhu and Lu, Bo-Wei and Yang, Chao-Wei and Liu, Run-Ze and Qiao, Yu-Kun and Li, Cheng and He, Zhi-Gang and Huang, Jia and Li, Hao and You, Li-Xing and Huo, Yong-Heng and Bao, Xiao-Hui and Pan, Jian-Wei},
    number = {16},
    month = {4},
    pages = {160801},
    volume = {132},
    publisher = {American Physical Society},
    doi = {10.1103/PhysRevLett.132.160801},
    issn = {0031-9007}
}

@article{Heupel2020,
    title = {{Fabrication and Characterization of Single-Crystal Diamond Membranes for Quantum Photonics with Tunable Microcavities}},
    year = {2020},
    journal = {Micromachines},
    author = {Heupel, Julia and Pallmann, Maximilian and K{\"{o}}rber, Jonathan and Merz, Rolf and Kopnarski, Michael and St{\"{o}}hr, Rainer and Reithmaier, Johann Peter and Hunger, David and Popov, Cyril},
    number = {12},
    month = {12},
    pages = {1080},
    volume = {11},
    publisher = {Multidisciplinary Digital Publishing Institute},
    doi = {10.3390/MI11121080},
    issn = {2072-666X}
}

@article{Babin2022,
    title = {{Fabrication and nanophotonic waveguide integration of silicon carbide colour centres with preserved spin-optical coherence}},
    year = {2022},
    journal = {Nat. Mater.},
    author = {Babin, Charles and St{\"{o}}hr, Rainer and Morioka, Naoya and Linkewitz, Tobias and Steidl, Timo and W{\"{o}}rnle, Raphael and Liu, Di and Hesselmeier, Erik and Vorobyov, Vadim and Denisenko, Andrej and Hentschel, Mario and Gobert, Christian and Berwian, Patrick and Astakhov, Georgy V. and Knolle, Wolfgang and Majety, Sridhar and Saha, Pranta and Radulaski, Marina and Son, Nguyen Tien and Ul-Hassan, Jawad and Kaiser, Florian and Wrachtrup, Jörg},
    month = {11},
    pages = {67--73},
    volume = {21},
    publisher = {Nature Publishing Group},
    doi = {10.1038/s41563-021-01148-3},
    issn = {1476-4660},
    pmid = {34795400}
}

@article{Janitz2015,
    title = {{Fabry-Perot microcavity for diamond-based photonics}},
    year = {2015},
    journal = {Phys. Rev. A},
    author = {Janitz, Erika and Ruf, Maximilian and Dimock, Mark and Bourassa, Alexandre and Sankey, Jack and Childress, Lilian},
    number = {4},
    month = {10},
    pages = {43844},
    volume = {92},
    doi = {10.1103/PhysRevA.92.043844},
    issn = {1050-2947, 1094-1622},
    language = {en}
}

@article{Koerber2024,
    title = {{Fluorescence Enhancement of Single V2 Centers in a 4H-SiC Cavity Antenna}},
    year = {2024},
    journal = {Nano Lett.},
    author = {K{\"{o}}rber, Jonathan and Heiler, Jonah and Fuchs, Philipp and Flad, Philipp and Hesselmeier, Erik and Kuna, Pierre and Ul-Hassan, Jawad and Knolle, Wolfgang and Becher, Christoph and Kaiser, Florian and Wrachtrup, Jörg},
    month = {7},
    pages = {53},
    volume = {13},
    publisher = {American Chemical Society},
    doi = {10.1021/acs.nanolett.4c02162},
    issn = {15306992},
    arxivId = {2406.08208}
}

@article{Hesselmeier2024,
    title = {{High fidelity optical readout of a nuclear spin qubit in Silicon Carbide}},
    year = {2024},
    journal = {Phys. Rev. Lett.},
    author = {Hesselmeier, Erik and Kuna, Pierre and Knolle, Wolfgang and Kaiser, Florian and Son, Nguyen Tien and Ghezellou, Misagh and Ul-Hassan, Jawad and Vorobyov, Vadim and Wrachtrup, Jörg},
    number = {18},
    month = {1},
    pages = {180804},
    volume = {132},
    publisher = {American Physical Society},
    doi = {10.1103/PhysRevLett.132.180804},
    issn = {10797114},
    arxivId = {2401.04465}
}

@article{Nagy2019,
    title = {{High-fidelity spin and optical control of single silicon-vacancy centres in silicon carbide}},
    year = {2019},
    journal = {Nat. Commun.},
    author = {Nagy, Roland and Niethammer, Matthias and Widmann, Matthias and Chen, Yu Chen and Udvarhelyi, Péter and Bonato, Cristian and Hassan, Jawad Ul and Karhu, Robin and Ivanov, Ivan G. and Son, Nguyen Tien and Maze, Jeronimo R. and Ohshima, Takeshi and Soykal, Oney O. and Gali, Ádám and Lee, Sang Yun and Kaiser, Florian and Wrachtrup, Jörg},
    number = {1},
    month = {4},
    pages = {1--8},
    volume = {10},
    publisher = {Nature Publishing Group},
    doi = {10.1038/s41467-019-09873-9},
    issn = {2041-1723},
    pmid = {31028260},
    arxivId = {1810.10296}
}

@article{Lukin2020,
    title = {{Integrated Quantum Photonics with Silicon Carbide: Challenges and Prospects}},
    shorttitle = {Integrated Quantum Photonics with Silicon Carbide},
    year = {2020},
    journal = {PRX Quantum},
    author = {Lukin, Daniil M and Guidry, Melissa A and Vu{\v{c}}kovi{\'{c}}, Jelena},
    number = {2},
    month = {12},
    pages = {020102},
    volume = {1},
    doi = {10.1103/PRXQuantum.1.020102},
    issn = {2691-3399},
    language = {en}
}

@article{Maisch2024,
    title = {{Investigation of Purcell enhancement of quantum dots emitting in the telecom O-band with an open fiber-cavity}},
    year = {2024},
    journal = {Phys. Rev. B},
    author = {Maisch, Julian and Grammel, Jonas and Tran, Nam and Jetter, Michael and Portalupi, Simone L. and Hunger, David and Michler, Peter},
    number = {16},
    month = {3},
    pages = {165301},
    volume = {110},
    publisher = {American Physical Society},
    doi = {https://doi.org/10.1103/PhysRevB.110.165301},
    issn = {24699969},
    arxivId = {2403.10960}
}

@article{Christle2017,
    title = {{Isolated spin qubits in SiC with a high-fidelity infrared spin-to-photon interface}},
    year = {2017},
    journal = {Phys. Rev. X},
    author = {Christle, David J. and Klimov, Paul V. and de las Casas, Charles F. and Sz{\'{a}}sz, Krisztián and Iv{\'{a}}dy, Viktor and Jokubavicius, Valdas and Hassan, Jawad Ul and Syv{\"{a}}j{\"{a}}rvi, Mikael and Koehl, William F. and Ohshima, Takeshi and Son, Nguyen T. and Janz{\'{e}}n, Erik and Gali, ádám and Awschalom, D.},
    number = {2},
    month = {6},
    pages = {021046},
    volume = {7},
    publisher = {American Physical Society},
    doi = {10.1103/PhysRevX.7.021046},
    issn = {21603308}
}

@article{Shang2020,
    title = {{Local vibrational modes of Si vacancy spin qubits in SiC}},
    year = {2020},
    journal = {Phys. Rev. B},
    author = {Shang, Z. and Hashemi, A. and Berenc{\'{e}}n, Y. and Komsa, H. P. and Erhart, P. and Zhou, S. and Helm, M. and Krasheninnikov, A. V. and Astakhov, G. V.},
    number = {14},
    month = {4},
    pages = {144109},
    volume = {101},
    publisher = {American Physical Society},
    doi = {10.1103/PhysRevB.101.144109},
    issn = {24699969},
    arxivId = {2002.00067}
}

@article{Hensen2015,
    title = {{Loophole-free Bell inequality violation using electron spins separated by 1.3 kilometres}},
    year = {2015},
    journal = {Nature },
    author = {Hensen, B. and Bernien, H. and Drea{\'{u}}, A. E. and Reiserer, A. and Kalb, N. and Blok, M. S. and Ruitenberg, J. and Vermeulen, R. F.L. and Schouten, R. N. and Abell{\'{a}}n, C. and Amaya, W. and Pruneri, V. and Mitchell, M. W. and Markham, M. and Twitchen, D. J. and Elkouss, D. and Wehner, S. and Taminiau, T. H. and Hanson, R.},
    number = {7575},
    month = {10},
    pages = {682--686},
    volume = {526},
    publisher = {Nature Publishing Group},
    doi = {10.1038/nature15759},
    issn = {1476-4687}
}

@article{Atatuere2018,
    title = {{Material platforms for spin-based photonic quantum technologies}},
    year = {2018},
    journal = {Nat. Rev. Mater.},
    author = {Atat{\"{u}}re, Mete and Englund, Dirk and Vamivakas, Nick and Lee, Sang Yun and Wrachtrup, Jörg},
    month = {4},
    pages = {38--51},
    volume = {3},
    publisher = {Nature Publishing Group},
    doi = {10.1038/s41578-018-0008-9},
    issn = {2058-8437}
}

@article{Balasubramanian2008,
    title = {{Nanoscale imaging magnetometry with diamond spins under ambient conditions}},
    year = {2008},
    journal = {Nature},
    author = {Balasubramanian, Gopalakrishnan and Chan, I. Y. and Kolesov, Roman and Al-Hmoud, Mohannad and Tisler, Julia and Shin, Chang and Kim, Changdong and Wojcik, Aleksander and Hemmer, Philip R. and Krueger, Anke and Hanke, Tobias and Leitenstorfer, Alfred and Bratschitsch, Rudolf and Jelezko, Fedor and Wrachtrup, Jörg},
    number = {7213},
    month = {10},
    pages = {648--651},
    volume = {455},
    publisher = {Nature Publishing Group},
    doi = {10.1038/nature07278},
    issn = {14764687}
}

@article{Hassan2008,
    title = {{On-axis homoepitaxial growth on Si-face 4H–SiC substrates}},
    year = {2008},
    journal = {J. Cryst. Growth},
    author = {Hassan, J. and Bergman, J. P. and Henry, A. and Janz{\'{e}}n, E.},
    number = {20},
    month = {10},
    pages = {4424--4429},
    volume = {310},
    publisher = {North-Holland},
    doi = {10.1016/j.jcrysgro.2008.06.081},
    issn = {0022-0248}
}

@article{Higginbottom2022,
    title = {{Optical observation of single spins in silicon}},
    year = {2022},
    journal = {Nature},
    author = {Higginbottom, Daniel B. and Kurkjian, Alexander T.K. and Chartrand, Camille and Kazemi, Moein and Brunelle, Nicholas A. and MacQuarrie, Evan R. and Klein, James R. and Lee-Hone, Nicholas R. and Stacho, Jakub and Ruether, Myles and Bowness, Camille and Bergeron, Laurent and DeAbreu, Adam and Harrigan, Stephen R. and Kanaganayagam, Joshua and Marsden, Danica W. and Richards, Timothy S. and Stott, Leea A. and Roorda, Sjoerd and Morse, Kevin J. and Thewalt, Michael L.W. and Simmons, Stephanie},
    number = {7918},
    month = {7},
    pages = {266--270},
    volume = {607},
    publisher = {Nature Publishing Group},
    doi = {10.1038/s41586-022-04821-y},
    issn = {1476-4687},
    pmid = {35831600},
    arxivId = {2103.07580}
}

@article{Nemoto2014,
    title = {{Photonic architecture for scalable quantum information processing in diamond}},
    year = {2014},
    journal = {Phys. Rev. X},
    author = {Nemoto, Kae and Trupke, Michael and Devitt, Simon J. and Stephens, Ashley M. and Scharfenberger, Burkhard and Buczak, Kathrin and N{\"{o}}bauer, Tobias and Everitt, Mark S. and Schmiedmayer, Jörg and Munro, William J.},
    number = {3},
    month = {8},
    pages = {031022},
    volume = {4},
    publisher = {American Physical Society},
    url = {https://journals.aps.org/prx/abstract/10.1103/PhysRevX.4.031022},
    doi = {10.1103/PhysRevX.4.031022},
    issn = {21603308},
    arxivId = {1309.4277}
}

@article{Zhou2023,
    title = {{Plasmonic-Enhanced Bright Single Spin Defects in Silicon Carbide Membranes}},
    year = {2023},
    journal = {Nano Lett.},
    author = {Zhou, Ji Yang and Li, Qiang and Hao, Zhi He and Lin, Wu Xi and He, Zhen Xuan and Liang, Rui Jian and Guo, Liping and Li, Hao and You, Lixing and Tang, Jian Shun and Xu, Jin Shi and Li, Chuan Feng and Guo, Guang Can},
    number = {10},
    month = {5},
    pages = {4334--4343},
    volume = {23},
    publisher = {American Chemical Society},
    doi = {10.1021/acs.nanolett.3c00568},
    issn = {15306992},
    pmid = {37155148}
}

@article{Krumrein2024,
    title = {{Precise Characterization of a Waveguide Fiber Interface in Silicon Carbide}},
    year = {2024},
    journal = {ACS Photonics},
    author = {Krumrein, Marcel and Nold, Raphael and Davidson-Marquis, Flavie and Bouamra, Arthur and Niechziol, Lukas and Steidl, Timo and Peng, Ruoming and K{\"{o}}rber, Jonathan and St{\"{o}}hr, Rainer and Gross, Nils and Smet, Jurgen H. and Ul-Hassan, Jawad and Udvarhelyi, Péter and Gali, Adam and Kaiser, Florian and Wrachtrup, Jörg},
    number = {6},
    month = {6},
    pages = {2160--2170},
    volume = {11},
    publisher = {American Chemical Society},
    url = {https://pubs.acs.org/doi/full/10.1021/acsphotonics.4c00538},
    doi = {10.1021/acsphotonics.4c00538},
    issn = {23304022}
}

@article{Crook2020,
    title = {{Purcell enhancement of a single silicon carbide color center with coherent spin control}},
    year = {2020},
    journal = {Nano Lett.},
    author = {Crook, Alexander L. and Anderson, Christopher P. and Miao, Kevin C. and Bourassa, Alexandre and Lee, Hope and Bayliss, Sam L. and Bracher, David O. and Zhang, Xingyu and Abe, Hiroshi and Ohshima, Takeshi and Hu, Evelyn L. and Awschalom, David D.},
    number = {5},
    month = {5},
    pages = {3427--3434},
    volume = {20},
    publisher = {American Chemical Society},
    doi = {10.1021/acs.nanolett.0c00339},
    issn = {15306992},
    pmid = {32208710},
    arxivId = {2003.00042}
}

@article{Awschalom2018,
    title = {{Quantum technologies with optically interfaced solid-state spins}},
    year = {2018},
    journal = {Nat. Photonics},
    author = {Awschalom, David D. and Hanson, Ronald and Wrachtrup, Jörg and Zhou, Brian B.},
    number = {9},
    month = {8},
    pages = {516--527},
    volume = {12},
    publisher = {Nature Publishing Group},
    doi = {10.1038/s41566-018-0232-2},
    issn = {1749-4893}
}

@article{Pompili2021,
    title = {{Realization of a multinode quantum network of remote solid-state qubits}},
    year = {2021},
    journal = {Science},
    author = {Pompili, M. and Hermans, S. L.N. and Baier, S. and Beukers, H. K.C. and Humphreys, P. C. and Schouten, R. N. and Vermeulen, R. F.L. and Tiggelman, M. J. and dos Santos Martins, L. and Dirkse, B. and Wehner, S. and Hanson, R.},
    number = {6539},
    month = {4},
    pages = {259--264},
    volume = {372},
    publisher = {American Association for the Advancement of Science},
    doi = {10.1126/science.abg1919},
    issn = {10959203},
    pmid = {33859028},
    arxivId = {2102.04471}
}

@article{Ruf2021,
    title = {{Resonant Excitation and Purcell Enhancement of Coherent Nitrogen-Vacancy Centers Coupled to a Fabry-Perot Microcavity}},
    year = {2021},
    journal = {Phys. Rev. Applied},
    author = {Ruf, M. and Weaver, M. J. and Van Dam, S. B. and Hanson, R.},
    number = {2},
    month = {2},
    pages = {024049},
    volume = {15},
    publisher = {American Physical Society},
    doi = {10.1103/PhysRevApplied.15.024049},
    issn = {23317019},
    arxivId = {2009.08204}
}

@article{Gaebel2006,
    title = {{Room-temperature coherent coupling of single spins in diamond}},
    year = {2006},
    journal = {Nat. Phys},
    author = {Gaebel, Torsten and Domhan, Michael and Popa, Iulian and Wittmann, Christoffer and Neumann, Philipp and Jelezko, Fedor and Rabeau, James R. and Stavrias, Nikolas and Greentree, Andrew D. and Prawer, Steven and Meijer, Jan and Twamley, Jason and Hemmer, Philip R. and Wrachtrup, Jörg},
    number = {6},
    pages = {408--413},
    volume = {2},
    publisher = {Nature Publishing Group},
    doi = {10.1038/nphys318},
    issn = {17452481}
}

@article{Li2022,
    title = {{Room-temperature coherent manipulation of single-spin qubits in silicon carbide with a high readout contrast}},
    year = {2022},
    journal = {Natl. Sci. Rev.},
    author = {Li, Qiang and Wang, Jun Feng and Yan, Fei Fei and Zhou, Ji Yang and Wang, Han Feng and Liu, He and Guo, Li Ping and Zhou, Xiong and Gali, Adam and Liu, Zheng Hao and Wang, Zu Qing and Sun, Kai and Guo, Guo Ping and Tang, Jian Shun and Li, Hao and You, Li Xing and Xu, Jin Shi and Li, Chuan Feng and Guo, Guang Can},
    number = {5},
    month = {6},
    volume = {9},
    publisher = {Oxford Academic},
    doi = {10.1093/nsr/nwab122},
    issn = {2095-5138},
    arxivId = {2005.07876}
}

@article{Kraus2013,
    title = {{Room-temperature quantum microwave emitters based on spin defects in silicon carbide}},
    year = {2014},
    journal = {Nat. Phys},
    author = {Kraus, H. and Soltamov, V. A. and Riedel, D. and V{\"{a}}th, S. and Fuchs, F. and Sperlich, A. and Baranov, P. G. and Dyakonov, V. and Astakhov, G. V.},
    number = {2},
    month = {12},
    pages = {157--162},
    volume = {10},
    publisher = {Nature Publishing Group},
    doi = {10.1038/nphys2826},
    issn = {1745-2481}
}

@article{Bekker2023,
    title = {{Scalable fabrication of hemispherical solid immersion lenses in silicon carbide through grayscale hard-mask lithography}},
    year = {2023},
    journal = {Appl. Phys. Lett.},
    author = {Bekker, Christiaan and Arshad, Muhammad Junaid and Cilibrizzi, Pasquale and Nikolatos, Charalampos and Lomax, Peter and Wood, Graham S. and Cheung, Rebecca and Knolle, Wolfgang and Ross, Neil and Gerardot, Brian and Bonato, Cristian},
    number = {17},
    month = {4},
    pages = {173507},
    volume = {122},
    publisher = {American Institute of Physics Inc.},
    doi = {doi.org/10.1063/5.0144684},
    issn = {00036951},
    arxivId = {2301.07705}
}

@article{Sardi2020,
    title = {{Scalable production of solid-immersion lenses for quantum emitters in silicon carbide}},
    year = {2020},
    journal = {Appl. Phys. Lett.},
    author = {Sardi, F. and Kornher, T. and Widmann, M. and Kolesov, R. and Schiller, F. and Reindl, T. and Hagel, M. and Wrachtrup, J.},
    number = {2},
    month = {7},
    pages = {22105},
    volume = {117},
    publisher = {American Institute of Physics Inc.},
    doi = {10.1063/5.0011366/39424},
    issn = {00036951}
}

@article{Radulaski2017,
    title = {{Scalable Quantum Photonics with Single Color Centers in Silicon Carbide}},
    year = {2017},
    journal = {Nano Lett.},
    author = {Radulaski, Marina and Widmann, Matthias and Niethammer, Matthias and Zhang, Jingyuan Linda and Lee, Sang-Yun and Rendler, Torsten and Lagoudakis, Konstantinos G. and Son, Nguyen Tien and Janz{\'{e}}n, Erik and Ohshima, Takeshi and Wrachtrup, Jörg and Vu{\v{c}}kovi{\'{c}}, Jelena},
    number = {3},
    month = {3},
    pages = {1782--1786},
    volume = {17},
    doi = {10.1021/acs.nanolett.6b05102},
    issn = {1530-6984}
}

@article{Koerber2023,
    title = {{Scanning Cavity Microscopy of a Single-Crystal Diamond Membrane}},
    year = {2023},
    journal = {Phys. Rev. Applied},
    author = {K{\"{o}}rber, Jonathan and Pallmann, Maximilian and Heupel, Julia and St{\"{o}}hr, Rainer and Vasilenko, Evgenij and H{\"{u}}mmer, Thomas and Kohler, Larissa and Popov, Cyril and Hunger, David},
    number = {6},
    month = {6},
    pages = {64057},
    volume = {19},
    doi = {10.1103/PhysRevApplied.19.064057},
    issn = {2331-7019},
    language = {en}
}

@article{Castelletto2020,
    title = {{Silicon carbide color centers for quantum applications}},
    year = {2020},
    journal = {JPhys Photonics},
    author = {Castelletto, Stefania and Boretti, Alberto and Phys Photonics, J and Moody, Galan and Sorger, Volker J and Blumenthal, Daniel J},
    number = {2},
    month = {3},
    pages = {022001},
    volume = {2},
    publisher = {IOP Publishing},
    doi = {10.1088/2515-7647/AB77A2},
    issn = {2515-7647}
}

@article{Castelletto2022,
    title = {{Silicon Carbide Photonics Bridging Quantum Technology}},
    year = {2022},
    journal = {ACS Photonics},
    author = {Castelletto, Stefania and Peruzzo, Alberto and Bonato, Cristian and Johnson, Brett C. and Radulaski, Marina and Ou, Haiyan and Kaiser, Florian and Wrachtrup, Jörg},
    number = {5},
    month = {5},
    pages = {1434--1457},
    volume = {9},
    publisher = {American Chemical Society},
    doi = {10.1021/acsphotonics.1c01775},
    issn = {23304022}
}

@article{Liu2015,
    title = {{Silicon carbide: A unique platform for metal-oxide-semiconductor physics}},
    year = {2015},
    journal = {Appl. Phys. Rev.},
    author = {Liu, Gang and Tuttle, Blair R. and Dhar, Sarit},
    number = {2},
    month = {6},
    pages = {021307},
    volume = {2},
    publisher = {American Institute of Physics Inc.},
    doi = {10.1063/1.4922748},
    issn = {19319401}
}

@article{Kang2020,
    title = {{Single-crystalline SiC integrated onto Si-based substrates via plasma-activated direct bonding}},
    year = {2020},
    journal = {Ceram. Int.},
    author = {Kang, Qiushi and Wang, Chenxi and Niu, Fanfan and Zhou, Shicheng and Xu, Jikai and Tian, Yanhong},
    number = {14},
    month = {10},
    pages = {22718--22726},
    volume = {46},
    publisher = {Elsevier},
    doi = {10.1016/j.ceramint.2020.06.036},
    issn = {0272-8842}
}

@article{Lai2024,
    title = {{Single-Shot Readout of a Nuclear Spin in Silicon Carbide}},
    year = {2024},
    journal = {Phys. Rev. Lett.},
    author = {Lai, Xiao-Yi and Fang, Ren-Zhou and Li, Tao and Su, Ren-Zhu and Huang, Jia and Li, Hao and You, Li-Xing and Bao, Xiao-Hui and Pan, Jian-Wei},
    number = {18},
    month = {1},
    pages = {180803},
    volume = {132},
    publisher = {American Physical Society},
    doi = {10.1103/PhysRevLett.132.180803},
    issn = {10797114},
    arxivId = {2401.04470}
}

@article{Neumann2010,
    title = {{Single-shot readout of a single nuclear spin}},
    year = {2010},
    journal = {Science},
    author = {Neumann, Philipp and Beck, Johannes and Steiner, Matthias and Rempp, Florian and Fedder, Helmut and Hemmer, Philip R. and Wrachtrup, Jörg and Jelezko, Fedor},
    number = {5991},
    month = {7},
    pages = {542--544},
    volume = {329},
    doi = {10.1126/science.1189075},
    issn = {00368075},
    pmid = {20595582}
}

@article{Heiler2024,
    title = {{Spectral stability of V2 centres in sub-micron 4H-SiC membranes}},
    year = {2024},
    journal = {npj quantum mater.},
    author = {Heiler, Jonah and K{\"{o}}rber, Jonathan and Hesselmeier, Erik and Kuna, Pierre and St{\"{o}}hr, Rainer and Fuchs, Philipp and Ghezellou, Misagh and Ul-Hassan, Jawad and Knolle, Wolfgang and Becher, Christoph and Kaiser, Florian and Wrachtrup, Jörg},
    number = {34},
    month = {4},
    volume = {9},
    publisher = {Nature Publishing Group},
    doi = {10.1038/s41535-024-00644-4},
    issn = {2397-4648}
}

@article{Kurtsiefer2001,
    title = {{The breakdown flash of silicon avalanche photodiodes-back door for eavesdropper attacks}},
    year = {2001},
    journal = {J. Mod. Opt.},
    author = {Kurtsiefer, Christian and Zarda, Patrick and Mayer, Sonja and Weinfurter, Harald},
    number = {13},
    month = {11},
    pages = {2039--2047},
    volume = {48},
    doi = {10.1080/09500340108240905},
    issn = {0950-0340}
}

@article{Liu2024,
    title = {{The silicon vacancy centers in SiC: determination of intrinsic spin dynamics for integrated quantum photonics}},
    year = {2024},
    journal = {npj quantum inf.},
    author = {Liu, Di and Kaiser, Florian and Bushmakin, Vladislav and Hesselmeier, Erik and Steidl, Timo and Ohshima, Takeshi and Son, Nguyen Tien and Ul-Hassan, Jawad and Soykal, Öney O. and Wrachtrup, Jörg},
    number = {1},
    month = {7},
    pages = {1--9},
    volume = {10},
    publisher = {Nature Publishing Group},
    doi = {10.1038/s41534-024-00861-6},
    issn = {2056-6387},
    arxivId = {2307.13648}
}

@article{Lukin2023,
    title = {{Two-Emitter Multimode Cavity Quantum Electrodynamics in Thin-Film Silicon Carbide Photonics}},
    year = {2023},
    journal = {Phys. Rev. X},
    author = {Lukin, Daniil M and Guidry, Melissa A and Yang, Joshua and Ghezellou, Misagh and Deb Mishra, Sattwik and Abe, Hiroshi and Ohshima, Takeshi and Ul-Hassan, Jawad and Vu{\v{c}}kovi{\'{c}}, Jelena},
    number = {1},
    month = {1},
    pages = {011005},
    volume = {13},
    doi = {10.1103/PhysRevX.13.011005},
    issn = {2160-3308},
    language = {en}
}

@article{Udvarhelyi2020,
    title = {{Vibronic States and Their Effect on the Temperature and Strain Dependence of Silicon-Vacancy Qubits in 4H-SiC}},
    year = {2020},
    journal = {Phys. Rev. Appl.},
    author = {Udvarhelyi, Péter and Thiering, Gergo and Morioka, Naoya and Babin, Charles and Kaiser, Florian and Lukin, Daniil and Ohshima, Takeshi and Ul-Hassan, Jawad and Son, Nguyen Tien and Vu{\v{c}}kovi{\'{c}}, Jelena and Wrachtrup, Jörg and Gali, Adam},
    number = {5},
    month = {5},
    pages = {054017},
    volume = {13},
    publisher = {American Physical Society},
    doi = {10.1103/PhysRevApplied.13.054017},
    issn = {23317019},
    arxivId = {2001.02459}
}

@article{Hollenbach2022,
    title = {{Wafer-scale nanofabrication of telecom single-photon emitters in silicon}},
    year = {2022},
    journal = {Nat. Commun.},
    author = {Hollenbach, Michael and Klingner, Nico and Jagtap, Nagesh S. and Bischoff, Lothar and Fowley, Ciarán and Kentsch, Ulrich and Hlawacek, Gregor and Erbe, Artur and Abrosimov, Nikolay V. and Helm, Manfred and Berenc{\'{e}}n, Yonder and Astakhov, Georgy V.},
    number = {1},
    month = {12},
    pages = {1--7},
    volume = {13},
    publisher = {Nature Publishing Group},
    url = {https://www.nature.com/articles/s41467-022-35051-5},
    doi = {10.1038/s41467-022-35051-5},
    issn = {2041-1723},
    pmid = {36509736},
    arxivId = {2204.13173}
}

@article{VanDam2018OptimalApproach,
    title = {{Optimal design of diamond-air microcavities for quantum networks using an analytical approach}},
    year = {2018},
    journal = {New. J. Phys.},
    author = {Van Dam, Suzanne B and Ruf, Maximilian and Hanson, Ronald},
    number = {11},
    month = {11},
    pages = {115004},
    volume = {20},
    doi = {10.1088/1367-2630/aaec29},
    issn = {1367-2630},
    language = {en}
}

@article{Fait2021,
    title = {{High finesse microcavities in the optical telecom O-band}},
    year = {2021},
    journal = {Applied Physics Letters},
    author = {Fait, J. and Putz, S. and Wachter, G. and Schalko, J. and Schmid, U. and Arndt, M. and Trupke, M.},
    number = {22},
    month = {11},
    pages = {14},
    volume = {119},
    publisher = {American Institute of Physics Inc.},
    url = {/aip/apl/article/119/22/221112/40930/High-finesse-microcavities-in-the-optical-telecom},
    doi = {10.1063/5.0066620/40930},
    issn = {00036951},
    arxivId = {2104.02813}
}

@article{Breev2021,
    title = {{Stress distribution at the AlN/SiC heterointerface probed by Raman spectroscopy}},
    year = {2021},
    journal = {Journal of Applied Physics},
    author = {Breev, I. D. and Likhachev, K. V. and Yakovleva, V. V. and H{\"{u}}bner, R. and Astakhov, G. V. and Baranov, P. G. and Mokhov, E. N. and Anisimov, A. N.},
    number = {5},
    month = {2},
    pages = {55304},
    volume = {129},
    publisher = {American Institute of Physics Inc.},
    url = {/aip/jap/article/129/5/055304/158142/Stress-distribution-at-the-AlN-SiC-heterointerface},
    doi = {10.1063/5.0029682/158142},
    issn = {10897550},
    arxivId = {2011.13693}
}
\end{document}


\title{Supplementary materials to: \\ Cavity enhancement of V2 centers in 4H-SiC with
a fiber-based Fabry-Pérot microcavity}

\author{Jannis Hessenauer}
\thanks{These authors contributed equally to this work}
\affiliation{
Physikalisches Institut, Karlsruhe Institute of Technology (KIT), Wolfgang-Gaede-Str. 1, 76131 Karlsruhe, Germany.
}

\author{Jonathan K\"orber}
\thanks{These authors contributed equally to this work}
\affiliation{
3rd Institute of Physics, University of Stuttgart, Pfaffenwaldring 57, 70569 Stuttgart, Germany.
}

\author{Misagh Ghezellou}
\affiliation{
Department of Physics, Chemistry and Biology, Linköping University, 581 83 Linköping, Sweden.
}

\author{Jawad Ul-Hassan}
\affiliation{
Department of Physics, Chemistry and Biology, Linköping University, 581 83 Linköping, Sweden.
}

\author{Georgy V. Astakhov}
\affiliation{
Institute of Ion Beam Physics and Materials Research, Helmholtz-Zentrum Dresden-Rossendorf, 01328 Dresden, Germany.
}

\author{Wolfgang Knolle}
\affiliation{
Leibniz-Institute of Surface Engineering (IOM), Permoserstraße 15, 04318 Leipzig, Germany.}

\author{J\"org Wrachtrup}
\affiliation{
3rd Institute of Physics, University of Stuttgart, Pfaffenwaldring 57, 70569 Stuttgart, Germany.
}
\affiliation{
Max Planck Institute for Solid State Research, Heisenbersgtraße 1, 70569 Stuttgart, Germany.}

\author{David Hunger}
\affiliation{
Physikalisches Institut, Karlsruhe Institute of Technology (KIT), Wolfgang-Gaede-Str. 1, 76131 Karlsruhe, Germany.
}
\affiliation{
Institute for Quantum Materials and Technologies (IQMT), Karlsruhe Institute of Technology (KIT), Herrmann-von-Helmholtz Platz 1, 76344 Eggenstein-Leopoldshafen, Germany.}

\date{\today} 
\maketitle

\section{Details on the sample fabrication} \label{sec:SuppFab}
\begin{figure}[htb]
\centering
   \includegraphics[width=1\textwidth]{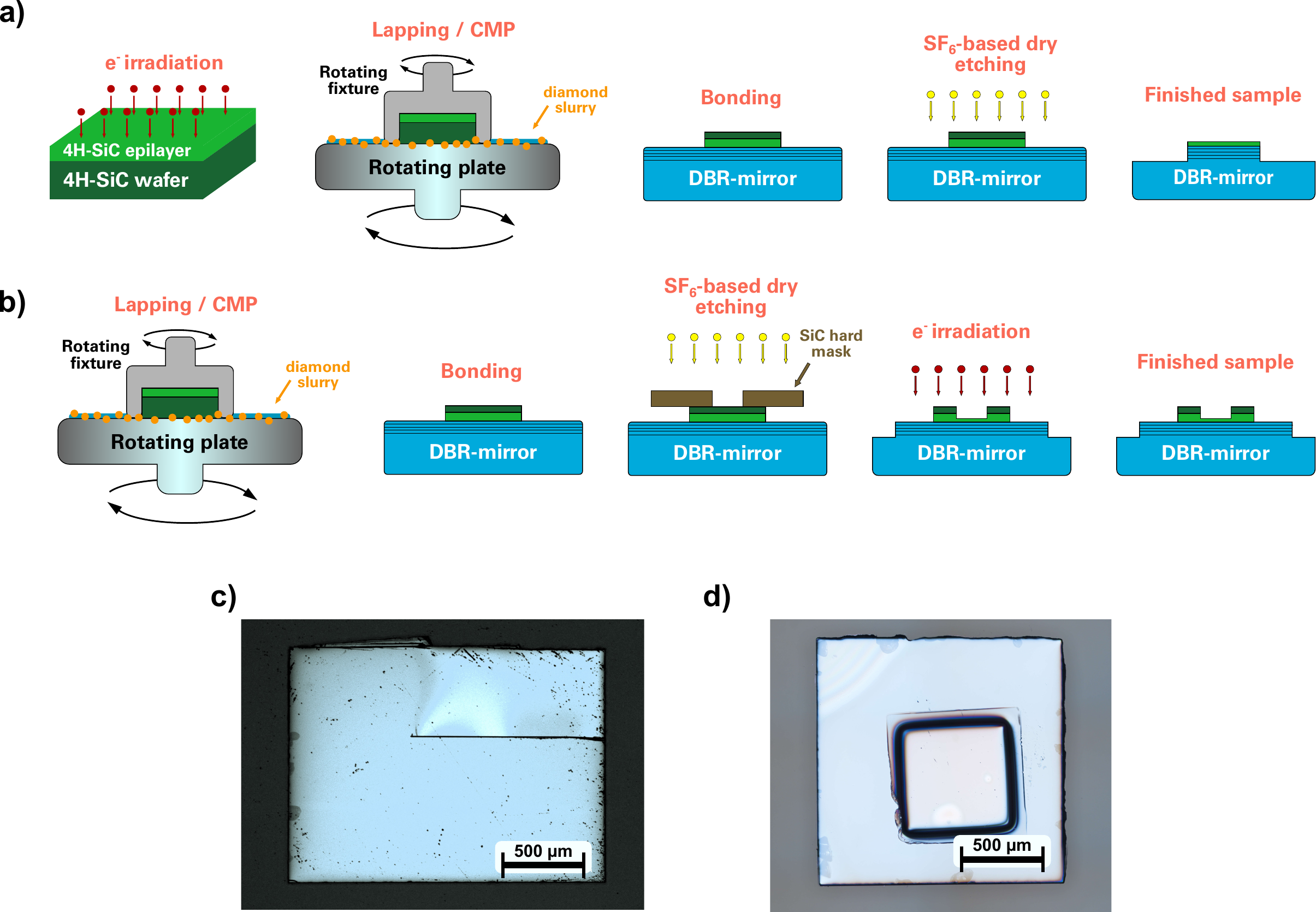}
\caption[]{Fabrication flow diagram of the two samples SA and SB. (a) Fabrication flow process of SB. (b) Fabrication flow process of SA. In contrast to (a) the electron irradiation was performed after the sample fabrication and a SiC hard mask was used during the dry etching, covering all parts apart from a central square opening. (c) and (d) Light microscope images of the samples SB and SA after the final etching. The use of a hard mask yields a membrane only within a square sub-region of SA (see black shaded square in (d)), while SB is fully thinned down. In comparison of both etched parts (membrane part of (d) and full region of (c)) one can see a clear difference in the density of etching pits (black dots).}
\label{SuppFig1}
\end{figure}
\subsection{Main samples of the study}
The fabrication of the samples SA and SB follows the process as depicted in figure \ref{SuppFig1} (b) and (a). For both samples of the experiments in the main manuscript (SA and SB), a commercial n-type a-plane 4H-SiC wafer (\textit{Wolfspeed}) is used. After a one-sided commercial chemical-mechanical polishing (\textit{NovaSiC}), n-type epilayers with a thickness of \SI{\sim 10}{\micro\meter} are grown along the a-axis on the polished side by chemical vapor deposition (CVD). For SA the isotope concentration is not controlled, leading to a natural abundance of isotopes in the epilayer. SB is taken from a different batch of epilayer growth, where the isotopes are controlled to \SI{0.5}{\percent} $^{13}$C and \SI{0.5}{\percent} $^{29}$Si in the CVD of the epilayer. We note that we do not expect significant differences from the varying isotope concentrations for our experiments at this stage, and we simply chose the two different epilayers on the basis of the availability of our samples. As stated in the main manuscript, we use a hard mask for the RIE process of SA, mainly to prevent micromasking during the etching, yielding only a small membrane part of the sample. For SB we decide to etch without the hard mask to achieve a larger region at the desired thickness. As can be seen in the microscope images of both samples after fabrication in figure \ref{SuppFig1} (c) and (d), the absence of a hard mask during etching leads to a higher density of etching marks for SB. However, the density of pits remains small enough compared to the lateral dimension of the cavity mode in our experiments, so that we are able to reach high-finesse values of \SI{\sim 40000}{} for both samples. As stated in the main text, we use electron irradiation for both samples to create V2-centers in the fabricated membranes. However, for SA we tried to create V2-centers using proton-implantation through a \SI{400}{\nano\meter}-thick PMMA mask with circular openings (diameters of \SIrange{200}{400}{\nano\meter}) in the first place. We used protons with an energy of \SI{27}{\kilo\electronvolt} and a dose of \SI{4e11}{\centi\meter^{-2}} for implantation. After we have not found V2 emission in the fiber cavity, we decided to additionally perform electron irradiation with this sample. In contrast, SB is electron irradiated already before fabrication. The electron irradiation is performed with an energy of \SI{5}{\mega\electronvolt} and a dose of \SI{5}{\kilo\gray} for both samples. The dose was determined by means of a \SI{5}{\milli\meter}-thick graphite calorimeter for the nominal energy of \SI{10}{\mega\electronvolt}. The nominal energy was reduced with a \SI{10}{\milli\meter}-thick aluminum plate to be about \SI{5}{\mega\electronvolt} onto the samples. The electron fluence can be calculated from the dose using the known average stopping power. For this, we use a conversion factor of $\SI{3.6}{\kilo\gray} \:\, \hat{=} \:\, \SI{1e13}{\centi\meter^{-2}}$. Thus, \SI{5}{\kilo\gray} correspond to a nominal fluence of \SI{\sim 1.4e13}{\centi\meter^{-2}}.
The use of the \SI{10}{\milli\meter}-thick aluminum plate leads to a loss of some lower-energy components of the electron beam and therefore to a slight reduction in the total electron fluence, which, however, is less than \SI{10}{\%}.

\subsection{Optically visible defects within sample B}
\begin{figure}[bt]
\centering
   \includegraphics[width=1\textwidth]{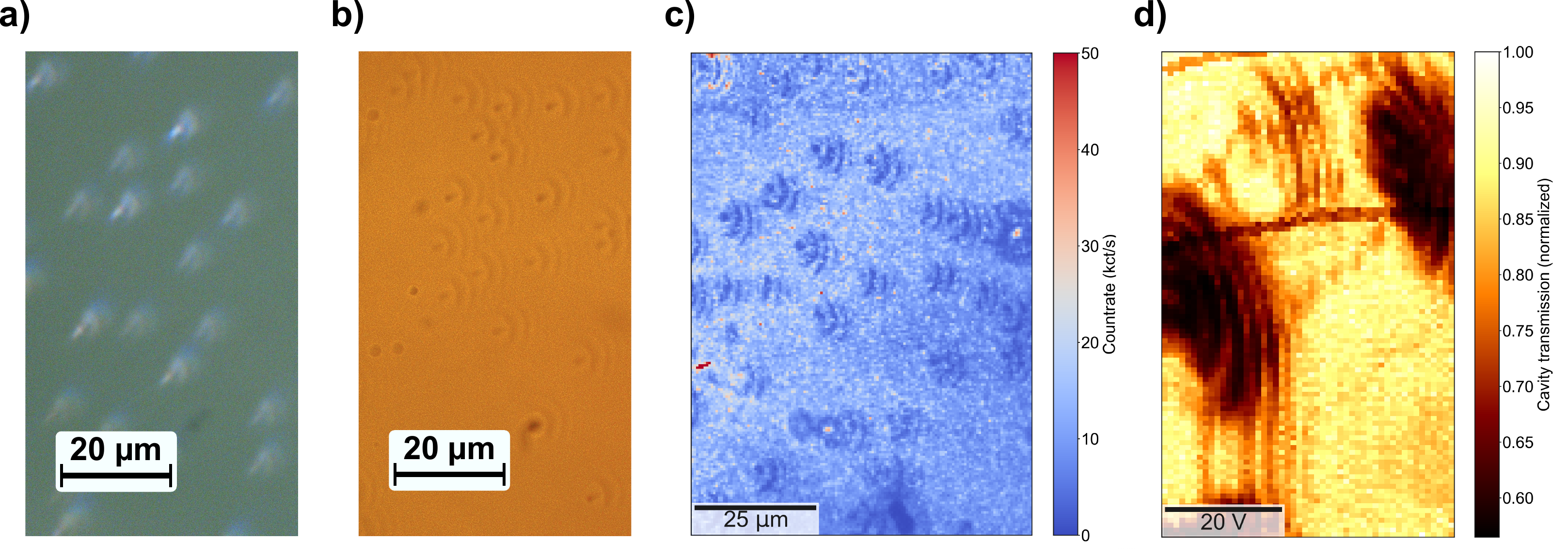}
\caption[]{
Optically visible defects in the SiC epilayer of SB.
Light microscope images of SB after bonding (a) and after the final etching (b). For image (a) the dark-field mode of the microscope is used, while image (b) is taken with the light-field mode. Both images show a high number of spots with a similar pattern. 
(c) Fluorescence map of SB taken in a room-temperature confocal microscope taken with \SI{850}{\micro\watt} of excitation power (\SI{785}{\nano\meter}). The defects are visible in a decrease of flurescence brightness.
(d) Cavity transmission scan of SB. The scale bar shows the voltage steps of the piezo, leading to the lateral cavity displacement.
}
\label{SuppFig2}
\end{figure}
During the fabrication of SB, we became aware of optically visible defects inside the epilayer of the SiC sample. As shown in Figs. \ref{SuppFig2} (a) and (b), these defects were already visible during the course of fabrication. After the fabrication and color center creation, we investigate the fluorescence of the sample in a room temperature confocal microscope and find those defects again, see Figure \ref{SuppFig2} (c). As expected, those defects lead to increased scattering of light in the fiber cavity experiments and thus are visible in a decreased cavity finesse and transmission, as depicted in figure figure \ref{SuppFig2} (d). We suspect that those defects originate from the epilayer growth. During the epitaxial growth process, a silicon-rich environment initially leads to the formation of Si droplets on the surface of the substrate at the start of the growth. However, as the growth processes, these droplets recrystallize into perfect SiC, leaving behind characteristic surface morphological features, as observed in the optical images of sample SB. Such defects are purely surface morphological in nature and do not represent any form of extended defects. Compared to the lateral mode extension of our fiber cavity, the density of the surface morphological features is small enough to find spots where the cavity can be operated without experiencing additional extinction.

\subsection{Van der Waals bonding}
\begin{figure}[bt]
\centering
   \includegraphics[width=1\textwidth]{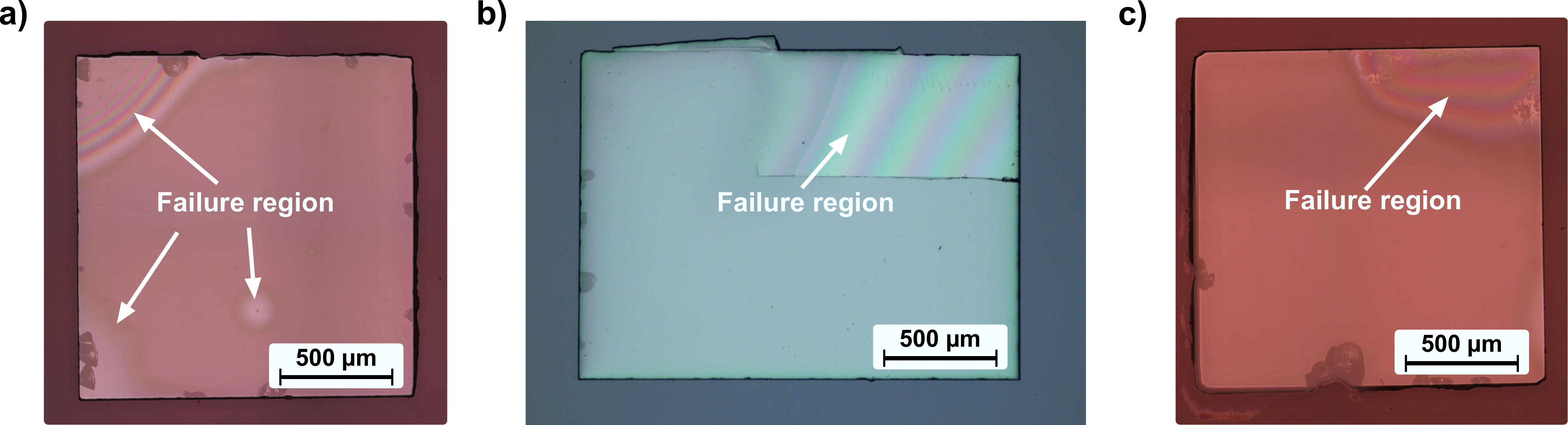}
\caption[]{
Light microscope images of the SiC samples after the Van der Waals bonding. Light microscope image of SA (a), SB (b), and a third sample that was bonded with the same procedure but used for a different study (c). All images were taken after the bond anneal. White arrows point towards regions where the contact bond failed which is visible through interference fringes, arising from the small airgap between the SiC and the mirror substrate.
}
\label{SuppFig3}
\end{figure}

In order to integrate the membrane samples into the fiber cavity, we bond them onto mirror substrates using Van der Waals forces, as described in the main manuscript. Once established, this results in a very strong bond between the SiC and the mirror enabling us to further process the samples, i.e. cleaning with organic solvents, dry etching and performing all the measurements in the cavity. To create the Van der Waals bond, we use the following recipe for our samples: First, the SiC samples are thoroughly cleaned by wiping cleanroom tissues dipped into acetone with strong pressure over the sample. This is repeated using isopropanol instead of acetone. The whole wiping is repeated until almost no remaining particles are visible on the surface under a light field microscope at high (50x) magnification. At next, the SiC and the mirror surface are activated using an oxygen plasma \cite{Kang2020}. For half of the samples we use an activation within our RIE-ICP machine (\textit{PlasmaPro80, Oxford Instruments}) with an RF and ICP power of \SI{100}{\watt}, a pressure of \SI{50}{\milli\torr}, an oxygen gas flow of \SI{30}{\sccm} and a time of 2 minutes. For the other half of our samples we use a softer treatment within a plasma cleaner (\textit{Pico plasmacleaner, diener}) at an ICP power of \SI{100}{\watt} a pressure of \SI{3}{\milli\torr}, an oxygen gas flow of \SI{30}{\sccm} and a time of 5 minutes, i.e. entirely without a directed plasma as it is used in the RIE-ICP. After the activation, we place each sample on top of a mirror substrate and apply manual pressure using the thumb and a small glass slide on top of the sample. Now, the sample should be already pre-bonded and not move anymore when applying pressure with tweezers from the side. After a successful pre-bond, we place the sample on a hot plate at maximum temperature of \SI{200}{^\circ \celcius} for several hours in order to strengthen the bond (bond anneal) \cite{Kang2020}. The results of the samples SA, SB and a further sample from a different study are shown in figure \ref{SuppFig3}. Apart from a few smaller regions, where the bond failed (visible by interference fringes stemming from the small airgap underneath), the samples are successfully Van der Waals bonded. With this recipe, we find a success rate of 2/3 and after repeating the process for the failed samples we are able to achieve a strong bond for all of our samples. Since we don't see differences in the success rate between both plasma activation steps, we decide to use the softer activation by the plasma cleaner for future experiments.

\subsection{AFM roughness after the fabrication}
\begin{figure}[tb]
\centering
   \includegraphics[width=0.9\textwidth]{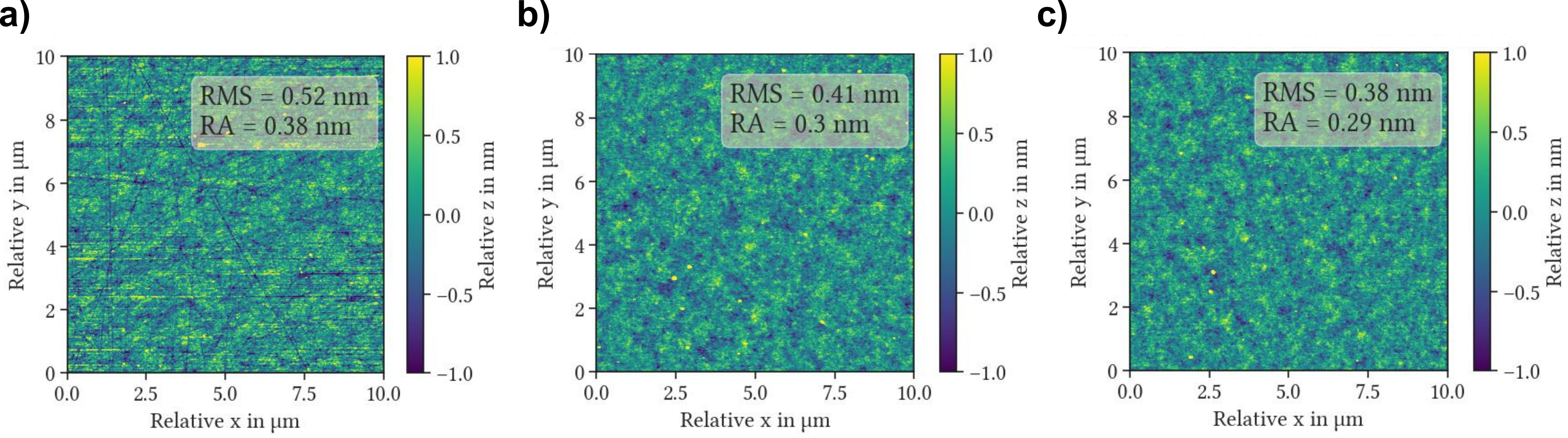}
\caption[]{
AFM measurements for surface roughness characterization of SA. AFM measurements of the surface of SA after the polishing (a) and for two different regions within the etched membrane (b) + (c). The RMS and RA deviation of the mean hight of the surface are shown on the upper right.}
\label{SuppFig4}
\end{figure}

As we have seen in previous work \cite{Heiler2024}, the surface RMS-roughness does not significantly change during the RIE process compared to the roughness after our polishing process. This is shown for SA with AFM measurements in figure \ref{SuppFig4}. Throughout the deep etching, the scratches that are left on the surface after the polishing (see figure \ref{SuppFig4} (a)) eventually vanish, leaving an even more homogeneous surface after the etching (see figure \ref{SuppFig4} (b) and (c)), which is beneficial for our cavity experiments.

\section{ODMR measurements for V2 identification at room temperature} \label{sec:SuppODMR}
\begin{figure}[tb]
\centering
   \includegraphics[width=1\textwidth]{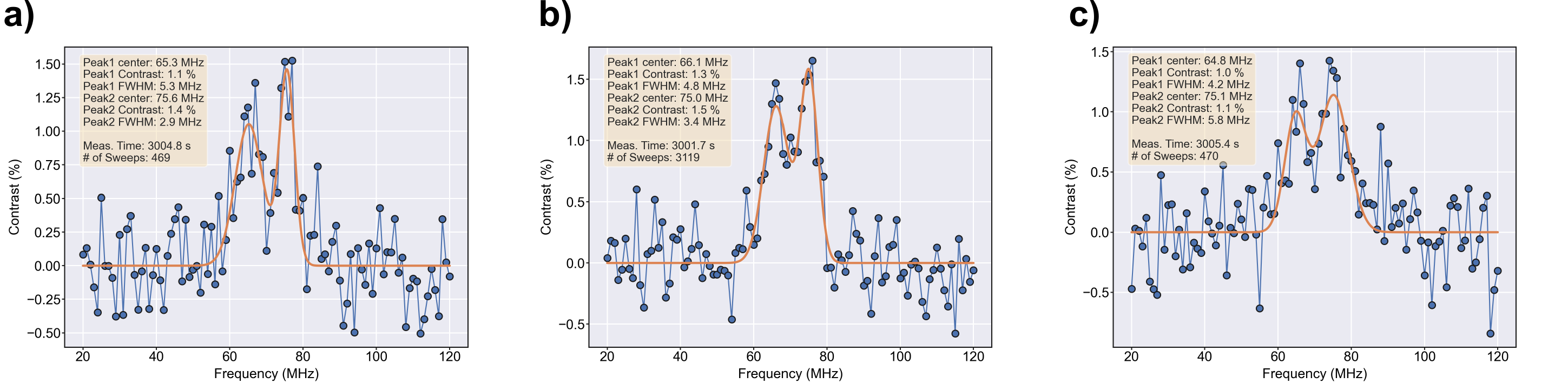}
\caption[]{
ODMR signal from several fluorescent spots of SA at room temperature. ODMR measurement at \SI{600}{\micro\watt} of optical excitation power and \SIrange{19}{20}{dbm} of microwave input power at three different spots on SA (a)-(c). Blue dots show the measurement results and the blue line is a guide to the eye. Solid, orange lines show a double-Lorentzian fit on the measured data.
}
\label{SuppFig5}
\end{figure}

As discussed in the main manuscript, we perform ODMR measurements in our room temperature confocal setup on SA to identify V2 centers among all fluorescent spots. To perform the measurements, we use a \SI{50}{\micro\meter}-thick copper wire spanned across the sample with a distance below \SI{100}{\micro\meter} to the investigated emitters. We send a microwave signal from a signal generator (\textit{SMIQ03, Rohde \& Schwarz}) after amplification (\textit{LZY-22+, Mini Circuit}) to \SIrange{19}{20}{dbm} through the wire while optically exciting with \SI{600}{\micro\watt} at \SI{785}{\nano\meter}. When sweeping the microwave frequency, we expect a positive signal in the fluorescence at \SI{70}{\mega\hertz} for V2 centers \cite{Kraus2013}. All spots with a clear ODMR contrast of \SI{\sim1}{\percent} or larger are counted as V2 centers in our statistics, as shown for three representative spots in figure \ref{SuppFig5}. We note that in our case the single ODMR peak is split into two peaks. This has been already seen in previous work with the setup \cite{Koerber2024} and most likely is a result of a magnetization of the objective.

\section{PLE measurements at different excitation power and temperature} \label{sec:SuppPLE}
\begin{figure}[tb]
\centering
   \includegraphics[width=1\textwidth]{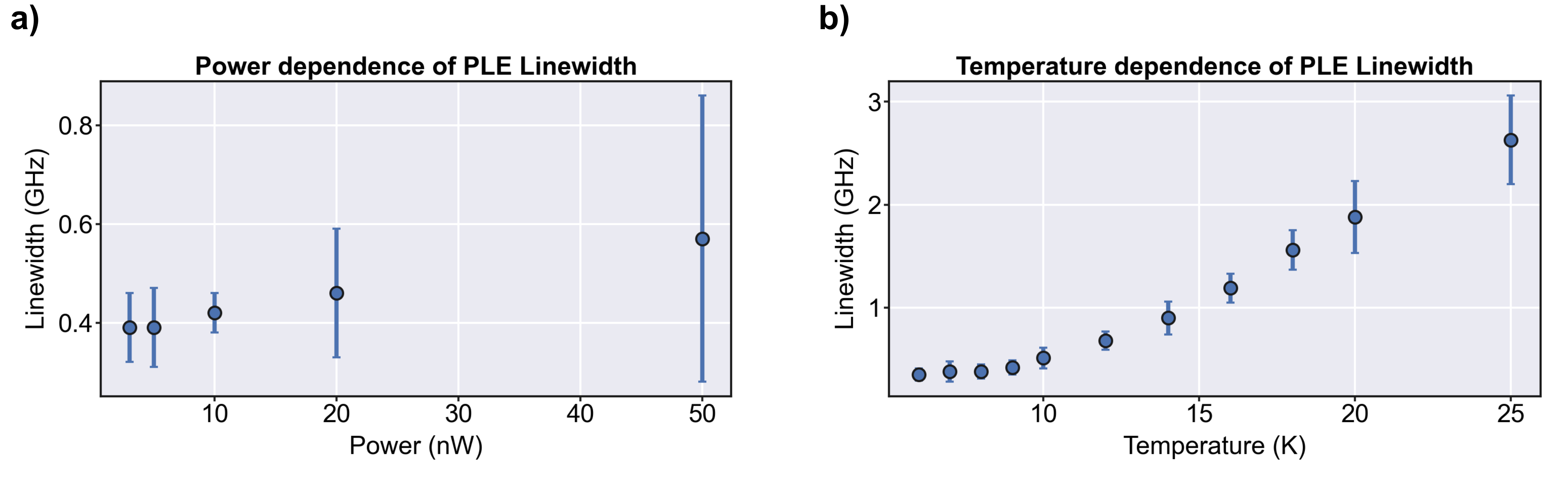}
\caption[]{
Excitation power and temperature dependence of PLE linewidth. Mean PLE linewidth for different excitation power at a set temperature of \SI{8}{\kelvin} (a) and different set temperature of the cryostation at a fixed excitation power of \SI{5}{\nano\watt} (b). The linewidths are the average of 25 single line PLE scans for each point. The errorbars indicate the standard deviation of each average linewidth.
}
\label{SuppFig6}
\end{figure}

To ensure that the PLE linewidths of our measurements on several emitters are not broadened due to a high excitation power or the temperature in the cryostation, we measure the PLE linewidth on a single V2 center as a function of the excitation power and the set cryostation temperature. To extract the PLE linewidth, we fit a Lorentzian on 25 single line scans for each excitation power (temperature) and extract the average and the standard deviation of all single lines. As shown in Figure \ref{SuppFig6} (a), the average linewidth starts to plateau at an excitation power of \SI{5}{\nano\watt}. Thus, we decide to measure PLE for all further spots with \SI{5}{\nano\watt} of excitation power. Similarly, the PLE linewidth plateaus at a set temperature below \SI{10}{\kelvin} of the cryostation, indicating that the measurements discussed in the main manuscript at a set temperature of \SI{8}{\kelvin} are not limited by temperature, as well. We note that the rather large standard deviation of the mean PLE linewidhts for higher excitation powers is a result of a higher number of ionization events within the 25 single line scans of our measurements. Since we do not exclude those lines from the average, the standard deviation becomes larger, while we are still able to extract the maximum excitation power without a broadening of the optical line.

\section{Mirror Coating}
\begin{figure}
    \centering
    \includegraphics[width=0.5\linewidth]{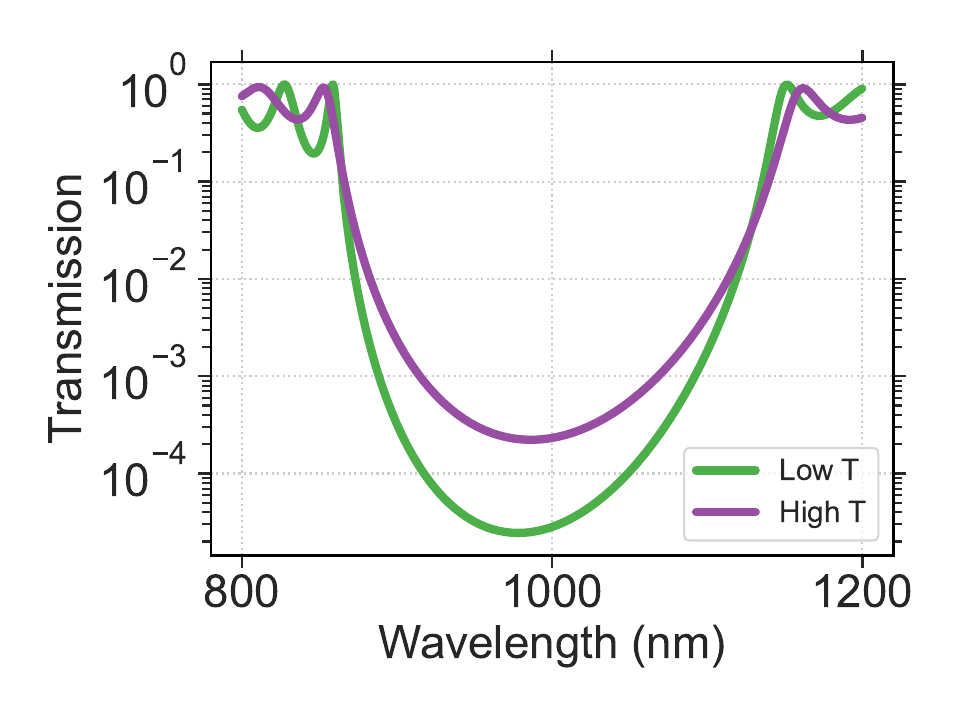}
    \caption{Stopband of the DBR mirrors used in this work. Calculated from the measured layer thickness provided by the coating manufacturer.}
    \label{fig:Stopband}
\end{figure}
Our planar and fiber mirrors are coated with a distributed Bragg reflector (DBR) layer stack, consisting of alternating $\lambda/4$ stacks of $\mathrm{SIO_2}$ and $\mathrm{Nb2O_5}$. The stopband is depicted in Figure \ref{fig:Stopband}. The center of the DBR stopband is nominally at $\SI{985}{nm}$, because the mirrors were initially designed for another experiment. We use an asymmetric cavity design, where the fiber mirror is a factor of $\approx 10$ times less transmissive than the planar mirror, in order to maximize the outcoupling efficiency towards the planar mirror. The nominal transmission of the planar mirror (fiber mirror) are $T_\mathrm{planar} = \SI{222}{ppm}$ and $T_\mathrm{fiber} = \SI{25}{ppm}$ at $\lambda=\SI{985}{nm}$. This results in a transmission limited finesse of $\mathcal{F} = \SI{25500}{}$, without considering absorption or scattering losses. The high transmission planar mirror is terminated with low refractive index layer and the high low transmission fiber mirror is terminated with a high refractive index layer. The membrane can act as an additional high refractive index stack and thereby improves the observable finesse in the air-like case. The nominal transmission values without a membrane at the ZPL wavelength of the V2 center $\lambda = \SI{917}{nm}$ are $T_\mathrm{planar} = \SI{900}{ppm}$ and $T_\mathrm{planar} = \SI{100}{ppm}$, resulting in a transmission limited finesse of $\mathcal{F}  \approx 6200 $.

\section{Cavity Setup}
\noindent We use a narrow-linewidth external cavity diode laser (Velocity TLB-6719, New Focus), tuneable from $\SIrange{940}{985}{nm}$, to probe the cavity properties such as the finesse. By blocking out the laserline, the amplified spontaneous emission background is also used as a broadband light source to characterize the dispersion. Additionally, a titanium sapphire laser (TiSa) (Solstis, MSquared) was used to  probe the cavity at the ZPL wavelength. Pulsed, offresonant exciation is provieded by a pulsed supercontinuum source (SuperK Fianium, NKT Photonics), providing broadband pulses with a pulsewidth $<\SI{100}{ps}$ and a repetition rate of up to $\SI{78}{MHz}$ .  We select a spectral window by means of an acousto-optic tunable filter (AOTF) (SuperK Select, NKT Photonics). We additionally use bandpass filters to suppress the broad background leaking through the AOTF. \\
Waveplates are used to adjust the polarization as needed before coupling the laser into optical fibers, which are connected as needed to the cavity fiber. \\
After the cavity, we collect and collimate the cavity mode transmitted through the planar mirror with a lense. The collected path is guided onto either
\begin{itemize}
    \item A sensitive photodiode (APD 410A, Thorlabs) to measure a narrow laser transmitted through the cavity. This is used for measuring the finesse or linewidth.
    \item A spectrometer (6 Shamrock 500i with iVAC316 LDC-DD camer, Andor Solis), used for measuring the cavity dispersion.
    \item Two single photon counters (Count NIR, LaserComponents) arranged in a Hanbury-Brown-Twiss (HBT) configuration. We use a polarizing beam splitter to split the light, allowing us to adjust the relative ratio depending on the measurement. This has the added benefit of partially suppressing the recombination background in the second order correlation function (see \ref{S:g2BG}). The pulses from the single photon counters are counted and analysed with correlation electronics (TimeTagger, Swabian Instruments).
    We filter out the $\SI{780}{nm}$ light using a long-pass  filter (FELH0800, Thorlabs). The HBT setup is shielded in a light-tight box and we use a BP filter (FBH 920-10, Thorlabs) at the entry aperture to suppress background light. In some measurements, we also added even narrower filtering using tunable short- and long-pass filters (Semrock), but found no significant difference.
\end{itemize}
The cavity setup has been reported on in \cite{Pallmann2023}. In short, we use motors for coarse movement of the planar mirror and piezo-lever mechanics to scan the fiber mirror in all three different spatial dimensions. The scanning voltages are provided by a digital-analogue converter (National Instruments) and are suitably amplified and filtered. 

\section{Recombination background in the g$^{(2)}$ measurement}
\label{S:g2BG}
\begin{figure}
    \centering
    \includegraphics[width=0.5\linewidth]{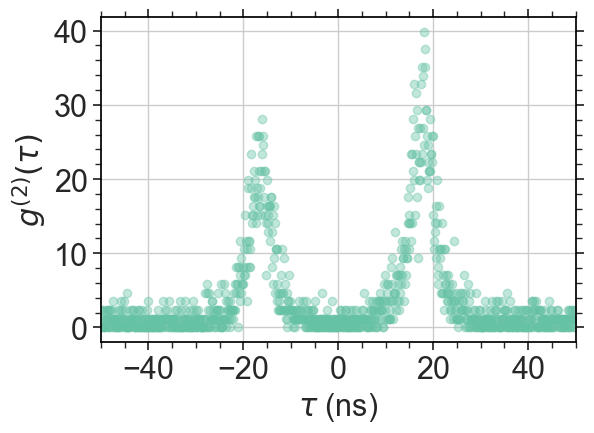}
    \caption{Characteristic background in the $g^{(2)}$ measurement. Data was taken with the cavity far detuned from any resonance under cw illumination, such that the overall countrate is given by the background.}
    \label{fig:devilshorns}
\end{figure}
When measuring second order autocorrelation functions, we always observe two characteristic peaks at time delay of $\SI{\pm 17}{ns}$, as depicted in Figure \ref{fig:devilshorns}  We attribute this to radiative recombination and relaxation processes in the silicon chips of the single photon counting modules, that are seen by the other detector, as previously observed \cite{Kurtsiefer2001}. The time delay corresponds to an optical path length of $d = \SI{5.2}{m}$, which is twice the distance between the cavity and the detectors. We observe these features regardless of the analysed light (Emitter and background fluorescence, narrow laser light or broadband light sources transmitted through the cavity, scattered flash-light radiation when illuminating only the detection path). Only the height of the features in the autocorrelation varies depending on the total countrate.
Notably, the peaks disappear when we illuminate the detectors while blocking the optical path from/to the cavity. We therefore attribute these peaks to backscattering of such recombination photons from the cavity. It is astonishing that the effect is so clearly visible, because only a small fraction of the photons emitted by this process should be guided in this specific optical mode.

Unfortunately, we can not spectrally filter this background, since it overlaps with the ZPL emission. Another approach that has been used previously was to utilize a combination of polarisers and polarizing beam splitter to isolate the detectors from each other \cite{Radulaski2017}. 
In fact, the polarizing beam-splitter used in the HBT here should already serve this role. But even with added polarization filters, we were unable to remove the artifacts completely. Therefore, we simply choose to exclude them from our fits. However, this precludes a more detailed analysis of the g$^{(2)}$ function that is useful to depend the rates governing the system, especially if done in a power dependent manner \cite{Pallmann2024}.  

\section{Cavity linewidth and stability}
\begin{figure*}[tb]
    \centering
    \includegraphics[width=0.75\linewidth]{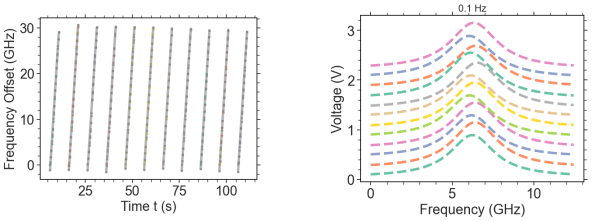}
    \caption{Measurement of the cavity linewidth. (a) To calibrate the cavity linewidth, wavemeter readings are recorded and fit with a linear slope. (b) 12 succinct fits to scans of the cavity line with 0.1 Hz, offset by 0.2 V each. The average linewidth is $\Delta \nu \approx \SI{3.44}{GHz}$.}
    \label{fig:cavity_lw}
\end{figure*}
\begin{figure*}[tb]
    \centering
    \includegraphics[width=0.8\linewidth]{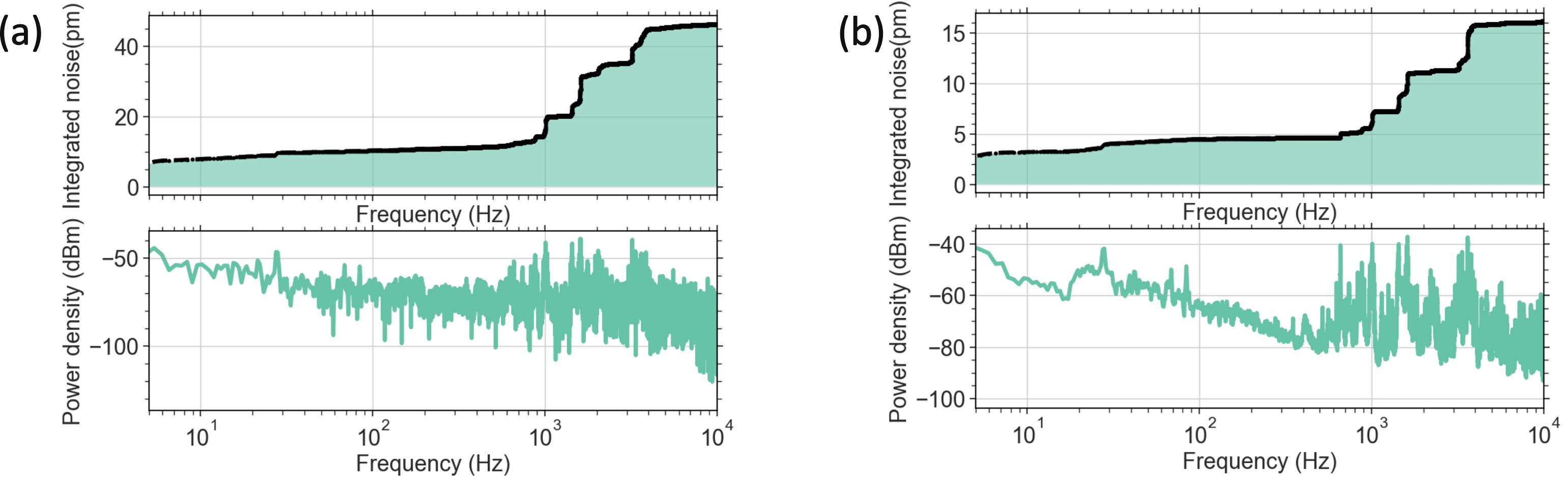}
    \caption{Cumulative RMS noise of the cavity and corresponding noise spectrum measured (a) for a free fiber mirror and (b) when the fiber mirror is brought in contact with the membrane-mirror system.}
    \label{fig:cavity_stability}    
\end{figure*}
\noindent We measure the cavity linewidth under the same conditions and at the same membrane position where we record the Purcell and $g^{(2)}(\tau)$ measurements depicted in Figure 4 of the main text. Therefore, we scan a laser tuned close to the emitter linewidth over the stationary cavity resonance and record the resonance. We calibrate the scan speed by fitting a linear slope to the recorded wavelength over time trace recorded with a wavemeter. We then fit the calibrated cavity line with a Lorentzian, yielding linewidths of $\Delta \nu \approx \SI{3.44}{GHz}$ for a scan speed of $\SI{6}{GHz/s}$ and $\Delta \nu \approx \SI{4.05}{GHz})$ for a scan speed of $\SI{0.6}{GHz/s}$. The difference is likely due to small drifts on long timescales. Exemplary linewidth  measurements are depicted in Figure \ref{fig:cavity_lw}. 

We also measure the cavity stability under cryogenic conditions by recording a fast fourier transform (FFT) of the cavity transmission while positioning the laser on the side of fringe of the cavity line. We can then calibrate the FFT amplitude in length fluctuations by the cavity finesse $\mathcal{F}$ \cite{Pallmann2023}. This results in $\sigma_\mathrm{RMS} \approx \SI{45}{pm}$ under normal operation and $\sigma_\mathrm{RMS} \approx \SI{16}{pm}$ when the fiber mirror is brought into contact with the membrane-mirror system (see Figure \ref{fig:cavity_stability}).

\section{Birefringence}
\begin{figure}[htbp]
    \centering
    \includegraphics[width=0.6\linewidth]{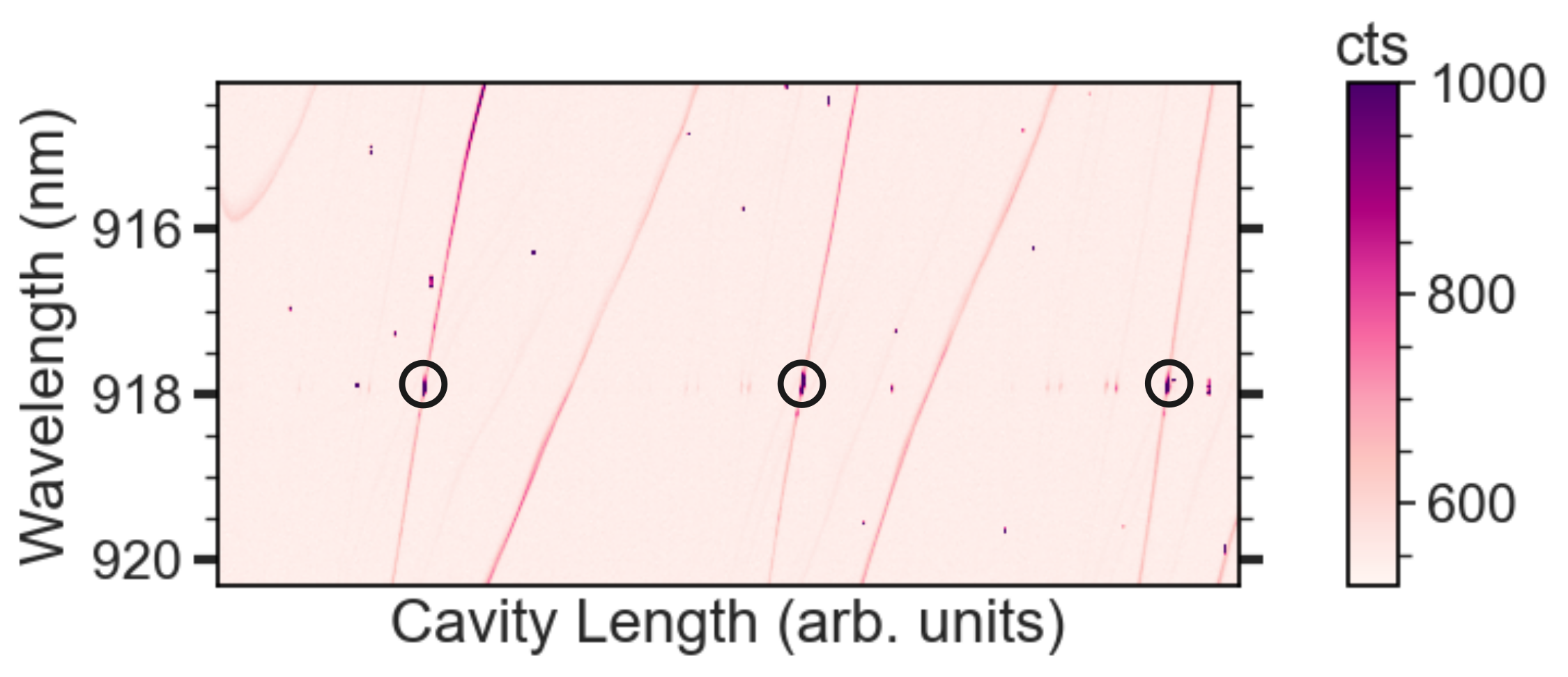}
    \caption{Cavity dispersion scan under off-resonant illumination. Only one set of modes couples to ZPLs (marked by black circles). }
    \label{fig:Dispersion}
\end{figure}

As mentioned in the main text, only one set of modes, corresponding to the extraordinary refractive index of the membrane, couples to the ZPLs of V2 centers. This is evident in a dispersion measurement under off resonant excitation at  $\lambda = \SI{785}{nm}$, as depicted in Figure \ref{fig:Dispersion}. The extraordinary and ordinary refractive index modes are distinguished by their different slope. It is evident that only one set of modes couples to the V2 center ZPL. 

%